\documentclass[sigconf,dvipsnames]{acmart}

\newif\ifcomments%

\commentsfalse%

\usepackage{ifthen}
\ifdefined\review%
	\ifthenelse{\equal{\review}{true}}{\commentstrue}{}
	\ifthenelse{\equal{\review}{false}}{\commentsfalse}{}
\fi

\ifcomments%
	\newlength{\marginNoteLength}
\fi

\usepackage{adjustbox}
\usepackage[ruled]{algorithm}
\usepackage{algorithmicx}
\usepackage[noend]{algpseudocode}
\usepackage{amsmath}
\usepackage{balance}
\usepackage[
	font=small,
	labelfont=bf
]{caption}
\usepackage{color}
\usepackage{colortbl}
\usepackage[
	lambda,
	advantage,
	operators,
	sets,
	landau,
	probability,
	notions,
	logic,
	ff,
	mm,
	primitives,
	events,
	complexity,
	asymptotics,
	keys
]{cryptocode}
\usepackage{dashbox}
\usepackage{diagbox}
\usepackage[inline]{enumitem}
\usepackage[mathscr]{eucal}
\usepackage{graphicx}
\usepackage{hyphenat}
\usepackage[utf8]{inputenc}
\usepackage{interval}
\usepackage{multicol}
\usepackage{multirow}
\usepackage{nicefrac}
\usepackage[binary-units]{siunitx}
\usepackage[capitalize]{cleveref}

\graphicspath{{./graphics/}}

\newcommand{\crypte}{Crypt$\epsilon$}

\newcommand{\record}{\ensuremath{r}}
\newcommand{\recordID}{\ensuremath{\record^\mathsf{ID}}}

\newcommand{\querySet}{\ensuremath{\mathcal{Q}}}
\newcommand{\queryKey}{\ensuremath{K}}

\newcommand{\domainSize}{\ensuremath{N}}
\newcommand{\dataSize}{\ensuremath{n}}
\newcommand{\oramsNumber}{\ensuremath{m}}

\newcommand{\user}{\ensuremath{\mathscr{U}}}
\newcommand{\server}{\ensuremath{\mathscr{S}}}

\newcommand{\protocol}{\ensuremath{\Pi}}
\newcommand{\protocolSetup}{\ensuremath{\protocol_{\mathsf{setup}}}}
\newcommand{\protocolQuery}{\ensuremath{\protocol_{\mathsf{query}}}}
\newcommand{\protocolNoGamma}{\ensuremath{\protocol_{\mathsf{no-}\gamma}}}
\newcommand{\protocolGamma}{\ensuremath{\protocol_\gamma}}

\newcommand{\searchKey}{\textsf{SK}}
\newcommand{\searchKeyDomain}{\ensuremath{\mathcal{X}}}

\newcommand{\serverDS}{\ensuremath{\mathcal{DS}}}
\newcommand{\indexI}{\ensuremath{\mathcal{I}}}
\newcommand{\database}{\ensuremath{\mathcal{D}}}
\newcommand{\databaseDef}{\ensuremath{\database = \allowbreak \{(\record_1, \allowbreak \recordID_1, \allowbreak \searchKey_1), \allowbreak \ldots, \allowbreak (\record_\dataSize, \allowbreak \recordID_\dataSize, \allowbreak \searchKey_\dataSize)\}}}
\newcommand{\fanout}{\ensuremath{k}}

\newcommand{\oram}{\ensuremath{\textsc{ORAM}}}
\newcommand{\oramProgram}{\ensuremath{\mathbf{y}}}
\newcommand{\oramRead}{\ensuremath{\mathbf{r}}}
\newcommand{\oramWrite}{\ensuremath{\mathbf{w}}}

\newcommand{\efficiencyCoefficient}{\ensuremath{a_1}}
\newcommand{\efficiencyOffset}{\ensuremath{a_2}}

\DeclareDocumentCommand{\algo}{ m g }{%
	{%
		\textsc{#1}%
		\IfNoValueF{#2}{\ensuremath{\left( #2 \right)}}%
	}%
}

\DeclareDocumentCommand{\query}{ g g }{%
	{%
		\IfValueTF{#2}%
			{\ensuremath{q_{\interval{#1}{#2}}}}%
			{
				\IfValueTF{#1}%
					{\ensuremath{q_{#1}}}%
					{\ensuremath{q}}
			}%
	}%
}

\newcommand{\adversary}{\ensuremath{\mathcal{A}}}

\renewcommand{\simulator}{\textsc{Sim}}
\DeclareDocumentCommand{\view}{ g g }{%
	{%
		\IfValueTF{#2}%
			{\ensuremath{\algo{View}_#1 \left( #2 \right)}}%
			{
				\IfValueTF{#1}%
					{\ensuremath{\algo{View}_#1}}%
					{\ensuremath{\algo{View}}}
			}%
	}%
}

\providecommand{\bigTheta}[1]{\ensuremath{\Theta \left( #1 \right)}}

\newcommand{\fromNtoM}[3]{\ensuremath{#1_#2, \ldots, #1_#3}}
\newcommand{\probability}[1]{\ensuremath{\textnormal{Pr}\left[ #1 \right]}} % chktex 35

\newcommand{\efficiency}[2]{\ensuremath{\left( \bigO{ #1 }, \ifthenelse{\equal{#2}{0}}{#2}{\bigO{ #2 }} \right)}}

\newtheorem{theorem}{Theorem}[section]

\newtheorem{corollary}[theorem]{Corollary}

\newtheorem{definition}[theorem]{Definition}
\newtheorem{remark}[theorem]{Remark}

\newcommand{\commentColor}[1]
{
	\ifthenelse{\equal{#1}{kellaris}}{\color{red}}{}
	\ifthenelse{\equal{#1}{kollios}}{\color{RubineRed}}{}
	\ifthenelse{\equal{#1}{dmytro}}{\color{blue}}{}
	\ifthenelse{\equal{#1}{adam}}{\color{ForestGreen}}{}
}

\makeatletter
	\newcommandx*{\sendmessageboth}[2][1=<->]{%
		\sendmessage{#1}{#2}%
	}
	\WithSuffix\newcommand\sendmessageboth*[2][\pcdefaultmessagelength]{%
		\begingroup%
			\renewcommand{\@pcsendmessagetop}{\let\halign\@pc@halign$\begin{aligned}#2\end{aligned}$}% chktex 21
			\sendmessage{<->}{length=#1}%
		\endgroup%
	}
\makeatother

\newcommand{\pcinput}[1]{\textbf{Input:}\ #1}
\newcommand{\pcouput}[1]{\textbf{Output:}\ #1}

\newcommand{\epsolute}{\ensuremath{\mathcal{E}}psolute}

\definecolor{urlColor}{RGB}{218,16,112}
\definecolor{lightGrey}{RGB}{220,220,220}

\makeatletter
	\let\orgdescriptionlabel\descriptionlabel%
	\renewcommand*{\descriptionlabel}[1]{%
		\let\orglabel\label%
		\let\label\@gobble% chktex 21
		\phantomsection%
		\protected@edef\@currentlabel{#1\unskip}% chktex 21
		\let\label\orglabel
		\orgdescriptionlabel{#1}%
	}
\makeatother

\AtBeginDocument{
	\pcfixhyperref%
	\makeatletter
		\crefformat{@pclinenumber}{line~#2#1#3}
		\Crefformat{@pclinenumber}{Line~#2#1#3}
		\crefrangeformat{@pclinenumber}{lines~#3#1#4 to~#5#2#6}
		\Crefrangeformat{@pclinenumber}{Lines~#3#1#4 to~#5#2#6}
	\makeatother
}

\hypersetup{
	colorlinks = true,
	citecolor = [RGB]{0, 87, 137},
	linkcolor = [RGB]{13, 33, 163},
	urlcolor = [RGB]{94, 155, 219}
}

\begin{document}

	% cSpell:ignore pdftitle pdfsubject pdfkeywords Johes SIGSAC

\copyrightyear{2021}
\acmYear{2021}
\setcopyright{acmlicensed}
\acmConference[CCS '21]{Proceedings of the 2021 ACM SIGSAC Conference on Computer and Communications Security}{November 15--19, 2021}{Virtual Event, Republic of Korea}
\acmBooktitle{Proceedings of the 2021 ACM SIGSAC Conference on Computer and Communications Security (CCS '21), November 15--19, 2021, Virtual Event, Republic of Korea}
\acmPrice{15.00}
\acmDOI{10.1145/3460120.3484786}
\acmISBN{978-1-4503-8454-4/21/11} % chktex 8
\settopmatter{printacmref=true}

\title{\texorpdfstring{\epsolute{}}{Epsolute}: Efficiently Querying Databases While Providing Differential Privacy}

\hypersetup{
	pdfinfo={
		Creator={\authors},
		Producer={\authors},
		Dedication={
			Dedicated to Dasha Bogatova.
			You are my love, my inspiration, my motivation, my support.
			I believe in you, there is nothing you cannot do!
		}
	}
}



\keywords{Differential Privacy; ORAM\@; differential obliviousness; sanitizers;}

\author{Dmytro Bogatov}
\affiliation{%
	\institution{Boston University}
	\city{Boston}
	\country{MA, USA}
}
\email{dmytro@bu.edu}

\author{Georgios Kellaris}
\affiliation{%
	\country{Canada}
}
\email{kellaris@bu.edu}

\author{George Kollios}
\affiliation{%
	\institution{Boston University}
	\city{Boston}
	\country{MA, USA}
}
\email{gkollios@cs.bu.edu}

\author{Kobbi Nissim}
\affiliation{%
	\institution{Georgetown University}
	\city{Washington}
	\country{D.C., USA}
}
\email{kobbi.nissim@georgetown.edu}

\author{Adam O'Neill}
\affiliation{%
	\institution{University of Massachusetts, Amherst}
	\city{Amherst}
	\country{MA, USA}
}
\email{adamo@cs.umass.edu}

	\begin{abstract}
		As organizations struggle with processing vast amounts of information, outsourcing sensitive data to third parties becomes a necessity.
To protect the data, various cryptographic techniques are used in outsourced database systems to ensure data privacy, while allowing efficient querying.
A rich collection of attacks on such systems has emerged.
Even with strong cryptography, just communication volume or access pattern is enough for an adversary to succeed.

In this work we present a model for \emph{differentially private outsourced database system} and a concrete construction, \epsolute{}, that provably conceals the aforementioned leakages, while remaining efficient and scalable.
In our solution, differential privacy is preserved at the record level even against an untrusted server that controls data and queries.
\epsolute{} combines Oblivious RAM and differentially private sanitizers to create a generic and efficient construction.

We go further and present a set of improvements to bring the solution to efficiency and practicality necessary for real-world adoption.
We describe the way to parallelize the operations, minimize the amount of noise, and reduce the number of network requests, while preserving the privacy guarantees.
We have run an extensive set of experiments, dozens of servers processing up to 10 million records, and compiled a detailed result analysis proving the efficiency and scalability of our solution.
While providing strong security and privacy guarantees we are less than an order of magnitude slower than range query execution of a non-secure plain-text optimized RDBMS like MySQL and PostgreSQL\@.

	\end{abstract}

	\maketitle

	% cSpell:ignore vely

\section{Introduction}

	Secure outsourced database systems aim at helping organizations outsource their data to untrusted third parties, without compromising data confidentiality or query efficiency.
	The main idea is to encrypt the data records before uploading them to an untrusted server along with an index data structure that governs which encrypted records to retrieve for each query.
	While strong cryptographic tools can be used for this task, existing implementations such as CryptDB~\cite{crypt-db}, Cipherbase~\cite{cipherbase}, StealthDB~\cite{stealth-db} and TrustedDB~\cite{trusted-db} try to optimize performance but do not provide strong security guarantees when answering queries.
	Indeed, a series of works~\cite{multidimensional-range-queries, inference-attack-islam-14, leakage-abuse-attacks-cash-15, inference-attacks-naveed-15, generic-attacks-kellaris, attacks-tao-of-inference, grubbs-attacks, access-pattern-disclosure, atacks-improved-reconstruction} demonstrate that these systems are vulnerable to a variety of reconstruction attacks.
	That is, an adversary can fully reconstruct the distribution of the records over the domain of the indexed attribute.
	This weakness is prominently due to the \emph{access pattern leakage}: the adversary can tell if the same encrypted record is returned on different queries.

	More recently, \cite{generic-attacks-kellaris, state-of-uniform, atacks-improved-reconstruction, pump-volume-attacks, volume-range-attacks} showed that reconstruction attacks are possible even if the systems employ heavyweight cryptographic techniques that hide the access patterns, such as homomorphic encryption~\cite{arbitrary-functions-encrypted, fully-homomorphic-encryption} or Oblivious RAM (ORAM)~\cite{oram-theory, oram-original}, because they leak the size of the result set of a query to the server (this is referred to as \emph{communication volume leakage}). % chktex 2
	Thus, even some recent systems that provide stronger security guarantees like ObliDB~\cite{oblidb}, Opaque~\cite{opaque} and Oblix~\cite{oblix} are susceptible to these attacks. % chktex 2
	This also means that no outsourced database system can be both optimally efficient and privacy-preserving: secure outsourced database systems should not return the exact number of records required to answer a query.

	We take the next step towards designing secure outsourced database systems by presenting novel constructions that strike a provable balance between efficiency and privacy.
	First, to combat the access pattern leakage, we integrate a layer of ORAM storage in our construction.
	Then, we bound the communication volume leakage by utilizing the notion of differential privacy (DP)~\cite{differential-privacy-original}.
	Specifically, instead of returning the exact number of records per query, we only reveal perturbed query answer sizes by adding random encrypted records to the result so that the communication volume leakage is bounded.
	Our construction guarantees privacy of any single record in the database which is necessary in datasets with stringent privacy requirements.
	In a medical HIPAA-compliant setting, for example, disclosing that a patient exists in a database with a rare diagnosis correlating with age may be enough to reveal a particular individual.

	The resulting mechanism achieves the required level of privacy, but implemented na\"{\i}vely the construction is prohibitively slow.
	We make the solution practical by limiting the amount of noise and the number of network roundtrips while preserving the privacy guarantees.
	We go further and present a way to parallelize the construction, which requires adapting noise-generation algorithms to maintain differential privacy requirements.

	Using our system, we have run an extensive set of experiments over cloud machines, utilizing large datasets --- that range up to 10 million records --- and queries of different sizes, and we report our experimental results on efficiency and scalability.
	We compare against best possible solutions in terms of efficiency (conventional non-secure outsourced database systems on unencrypted data) and against an approach that provides optimal security (retrieves the full table from the cloud or runs the entire query obliviously with maximal padding).
	We report that our solution is very competitive against both baselines.
	Our performance is comparable to that of unsecured plain-text optimized database systems (like MySQL and PostgreSQL): while providing strong security and privacy guarantees, we are only 4 to 8 times slower in a typical setting.
	Compared with the optimally secure solution, a linear scan (downloading all the records), we are 18 times faster in a typical setting and even faster as database sizes scale up.

	\smallskip

	To summarize, our contributions in this work are as follows:
	\begin{itemize}
		\item
			We present a new model for a \emph{differentially private} outsourced database system, CDP-ODB, its security definition, query types, and efficiency measures.
			In our model, the adversarial honest-but-curious server cannot see the record values, access patterns, or exact communication volume.

		\item
			We describe a novel construction, \epsolute{}, that satisfies the proposed security definition, and provide detailed algorithms for both range and point query types.
			In particular, to conceal the access pattern and communication volume leakages, we provide a secure storage construction, utilizing a combination of Oblivious RAM~\cite{oram-theory, oram-original} and differentially private sanitization~\cite{non-interactive-database-privacy}.
			Towards this, we maintain an index structure to know how many and which objects we need to retrieve.
			This index can be stored locally for better efficiency (in all our experiments this is the case), but crucially, it can also be outsourced to the adversarial server and retrieved on-the-fly for each query.

		\item
			We improve our generic construction to enable parallelization within a query.
			The core idea is to split the storage among multiple ORAMs, but this requires tailoring the overhead required for differential privacy proportionally to the number of ORAMs, in order to ensure privacy.
			We present practical improvements and optimization techniques that dramatically reduce the amount of fetched noise and the number of network roundtrips.

		\item
			Finally, we provide and open-source a high-quality C++ implementation of our system.
			We have run an extensive set of experiments on both synthetic and real datasets to empirically assess the efficiency of our construction and the impact of our improvements.
			We compare our solutions to the na\"{\i}ve approach (linear scan downloading all data every query), oblivious processing and maximal padding solution (Shrinkwrap~\cite{shrinkwrap}), and to a non-secure regular RDBMS (PostgreSQL and MySQL), and we show that our system is very competitive.
	\end{itemize}

	\subsection{Related Work}

		We group the related secure databases, engines, and indices into three categories
		\begin{enumerate*}[label={(\roman*)}]
			\item systems that are oblivious or volume-hiding and do not require trusted execution environment (TEE),
			\item constructions that rely on TEE (usually, Intel SGX),
			\item solutions that use property-preserving or semantically secure encryption and target primarily a snapshot adversary.
		\end{enumerate*}
		\emph{We claim that \epsolute{} is the most secure and practical range- and point-query engine in the outsourced database model, that protects both access pattern (AP) and communication volume (CV) using Differential Privacy, while not relying on TEE, linear scan or padding result size to the maximum.}

		\paragraph*{Obliviousness and volume-hiding without enclave}

			This category is the most relevant to \epsolute{}, wherein the systems provide either or both AP and CV protection without relying on TEE\@.
			\crypte{}~\cite{crypte} is a recent end-to-end system executing ``DP programs''.
			\crypte{} has a different model than \epsolute{} in that it assumes two non-colluding servers, an adversarial querying user (the analyst), and it uses DP to protect the privacy of an individual in the database, which includes volume-hiding for aggregate queries.
			\crypte{} also does not consider oblivious execution and attacks against the AP\@.
			Shrinkwrap~\cite{shrinkwrap} (and its predecessor SMCQL~\cite{smcql}) is an excellent system designed for complex queries over federated and distributed data sources.
			In Shrinkwrap, AP protection is achieved by using oblivious operators (linear scan and sort) and CV is concealed by adding fake records to intermediate results with DP\@.
			Padding the result to the maximum size first and doing a linear scan over it afterwards to ``shrink'' it using DP, is much more expensive than in \epsolute{}, however.
			In addition, in processing a query, the worker nodes are performing an $O(n \log{n})$ cost oblivious sorting, where $n$ is the maximum result size (whole table for range query), since they are designed to answer more general complex queries.
			SEAL~\cite{seal} offers adjustable AP and CV leakages, up to specific bits of leakage.
			SEAL builds on top of Logarithmic-SRC~\cite{practical-range-search}, splits storage into multiple ORAMs to adjust AP, and pads results size to a power of 2 to adjust CV\@.
			\epsolute{}, on the other hand, fully hides the AP and uses DP with its guarantees to pad the result size.
			PINED-RQ~\cite{pined-rq} samples Laplacian noise right in the B+ tree index tree, adding fake and removing real pointers according to the sample.
			Unlike \epsolute{}, PINED-RQ allows false negatives (i.e., result records not included in the answer), and does not protect against AP leakage.
			On the theoretical side, \citet{differential-obliviousness} (followed by \citet{differential-obliviousness-followup}) treat the AP itself as something to protect with DP\@.
			\cite{differential-obliviousness} introduces a notion of differential obliviousness that is admittedly weaker than the full obliviousness used in \epsolute{}. % chktex 2
			Most importantly, \cite{differential-obliviousness} ensures differential privacy w.r.t.~the ORAM only, while \epsolute{} ensures DP w.r.t.~the entire view of the adversary. % chktex 2

		\paragraph*{Enclave-based solutions}

			Works in this category use trusted execution environment (usually, SGX enclave).
			These works are primarily concerned with the AP protection for both trusted and untrusted memory, unlike \epsolute{} which also protects CV\@.
			Cipherbase~\cite{cipherbase,cipherbase-daas} was a pioneer introducing the idea of using TEE (FPGA at that time) to assist with DBMS security.
			HardIDX~\cite{hardidx} simply puts the B+ tree in the enclave, while StealthDB~\cite{stealth-db} symmetrically encrypts all records and brings them in the enclave one at a time for processing.
			EnclaveDB~\cite{enclave-db} assumes somewhat unrealistic \SI{192}{\giga\byte} enclave and puts the entire database in it.
			ObliDB~\cite{oblidb} and Opaque~\cite{opaque} assume fully oblivious enclave memory (not available as of today) and devise algorithms that use this fully trusted portion to obliviously execute common DBMS operators, like filters and joins.
			Oblix~\cite{oblix} provides a multimap that is oblivious both in and out of the enclave.
			HybrIDX claims protection against both AP and CV leakages, but unlike \epsolute{} it only obfuscates them.
			\epsolute{} offers an indistinguishability guarantee for AP and a DP guarantee for CV, while HybrIDX hides the exact result size and only obfuscates the AP\@.
			Lastly, Hermetic~\cite{hermetic} takes on the SGX side-channel attacks, including AP\@.
			It provides oblivious primitives, however, it only offers protection against software and not physical attacks (e.g., it trusts a hypervisor to disable interrupts).

		\paragraph*{Solutions against the snapshot adversary}

			Works in this category protect against the snapshot adversary, which takes a snapshot of the data at a fixed point in time (e.g., stolen hard drive).
			We stress that \epsolute{} provides semantic security against the snapshot adversary on top of AP and CV protection.
			CryptDB~\cite{crypt-db} is a seminal work in this direction offering computations over encrypted data.
			It has since been shown (e.g.~\cite{inference-attacks-naveed-15,inference-attack-islam-14,attacks-tao-of-inference}) that the underlying property-preserving schemes allow for reconstruction attacks.
			Arx~\cite{arx} provides strictly stronger security guarantees by using only semantically secure primitives.
			Seabed~\cite{seabed} uses an additively symmetric homomorphic encryption scheme for aggregates and certain filter queries.
			\citet{ppqed} offer a method to verify and apply a predicate (a junction of conditions) using garbled circuits or homomorphic encryption without revealing the predicate itself.
			SisoSPIR~\cite{sisospir} presents a mechanism to build an oblivious index tree such that neither party learns the pass taken.
			See \cite{ore-benchmark-17} for a survey of range query protocols in this category. % chktex 2

	\section{Background}\label{section:background}

	In this section we describe \emph{an outsourced database system} adapted from~\cite{generic-attacks-kellaris}, a base for our own model (\cref{section:dpodb}), and the constructions we will use as building blocks in our solution.

	\subsection{Outsourced Database System}

		We abstract a database as a collection of \dataSize{} records \record{}, each with a unique identifier \recordID{}, associated with search keys \searchKey{}: \databaseDef{}.
		We assume that all records have an identical fixed bit-length, and that search keys are elements of the domain $\searchKeyDomain = \{ 1, \ldots, \domainSize \}$ for some $\domainSize \in \NN$.
		Outsourced database systems support search keys on multiple attributes, with a set of search keys for each of the attributes of a record.
		For the ease of presentation, we describe the model for a single indexed attribute and then show how to extend it to support multiple attributes.

		A query is a predicate $\query: \searchKeyDomain \to \bin$.
		Evaluating a query \query{} on a database \database{} results in $\query( \database ) = \{ \record_i : \query( \searchKey_i ) = 1 \}$, all records whose search keys satisfy $\query$.

		Let \querySet{} be a set of queries.
		An \emph{outsourced database system} for queries in \querySet{} consists of two protocols between two \emph{stateful} parties: a user \user{} and a server \server{} (adapted from \cite{generic-attacks-kellaris}): % chktex 2
		\begin{description}
			\item[Setup protocol \protocolSetup{}:]
				\user{} receives as input a database \databaseDef{}; \server{} has no input.
				The output for \server{} is a data structure \serverDS{}; \user{} has no output besides its state.

			\item[Query protocol \protocolQuery{}:]
				\user{} has a query $\query \in \querySet$ produced in the setup protocol as input; \server{} has as input \serverDS{} produced in the setup protocol.
				\user{} outputs $\query( \database )$; \server{} has no formal output.
				(Both parties may update their internal states.)
		\end{description}

		For correctness, we require that for any database \databaseDef{} and query $\query \in \querySet$, it holds that running \protocolSetup{} and then \protocolQuery{} on the corresponding inputs yields for $\user{}$ the correct output $\{ \record_i : \query( \searchKey_i ) = 1 \}$ with overwhelming probability over the coins of the above runs.
		We call the protocol $\eta$-wrong if this probability is at least $1 - \eta$.

	\subsection{Differential Privacy and Sanitization}\label{section:building-blocks:dp}

		Differential privacy is a definition of privacy in analysis that protects information that is specific to individual records.
		More formally, we call databases $\database_1 \in \searchKeyDomain^\dataSize$ and $\database_2 \in \searchKeyDomain^\dataSize$ over domain \searchKeyDomain{} \emph{neighboring} (denoted $\database_1 \sim \database_2$) if they differ in exactly one record.

		\begin{definition}[\cite{our-data-ourselves, differential-privacy-original}]

			A randomized algorithm \algo{A} is $(\epsilon, \delta)$-differentially private if for all $\database_1 \sim \database_2 \in \searchKeyDomain^\dataSize$, and for all subsets $\mathcal{O}$ of the output space of \algo{A},
			\[
				\probability{ \algo{A}{ \database_1 } \in \mathcal{O} } \leq \exp(\epsilon) \cdot \probability{ \algo{A}{ \database_2 } \in \mathcal{O} } + \delta \; .
			\]
			The probability is taken over the random coins of \algo{A}.
		\end{definition}

		When $\delta = 0$ we omit it and say that \algo{A} preserves \emph{pure} differential privacy, otherwise (when $\delta > 0$) we say that \algo{A} preserves \emph{approximate} differential privacy.

		We will use mechanisms for answering count queries with differential privacy.
		Such mechanisms perturb their output to mask out the effect of any single record on their outcome.
		The simplest method for answering count queries with differential privacy is the Laplace Perturbation Algorithm (LPA)~\cite{differential-privacy-original} where random noise drawn from a Laplace distribution is added to the count to be published.
		The noise is scaled so as to hide the effect any single record can have on the count.
		More generally, the LPA can be used to approximate any statistical result by scaling the noise to the \emph{sensitivity} of the statistical analysis.\footnote{
			The \emph{sensitivity} of a query \query{} mapping databases into $\RR^\domainSize$ is defined to be $\Delta(\query) = \max_{\database_1 \sim \database_2 \in \searchKeyDomain^\dataSize} \norm{ \query(\database_1) - \query(\database_2) }_1$.
		}

		\begin{theorem}[adapted Theorem 1 from~\cite{differential-privacy-original}]\label{theorem:lpa}
			Let $\query : \database \to \RR^\domainSize$.
			An algorithm \algo{A} that adds independently generated noise from a zero-mean Laplace distribution with scale $\lambda = \nicefrac{\Delta(\query)}{\epsilon}$ to each of the \domainSize{} coordinates of $\query(\database)$, satisfies $\epsilon$-differential privacy.
		\end{theorem}

		While \cref{theorem:lpa} is an effective and simple way of answering a single count query, we will need to answer a sequence of count queries, ideally, without imposing a bound on the length of this sequence.
		We will hence make use of \emph{sanitization} algorithms.

		\begin{definition}\label{definition:dp-danitizer}
			Let \querySet{} be a collection of queries.
			An $(\epsilon, \delta, \alpha, \beta)$-differentially private sanitizer for \querySet{} is a pair of algorithms $(\algo{A}, \algo{B})$ such that:
			\begin{itemize}
				\item $A$ is $(\epsilon, \delta)$-differentially private, and
				\item on input a dataset $\database = \fromNtoM{d}{1}{\dataSize} \in \searchKeyDomain^\dataSize$, \algo{A} outputs a data structure \serverDS{} such that with probability $1 - \beta$ for all $\query \in \querySet$, $\abs{ \algo{B}{ \serverDS, \query } - \sum_i \query(d_i) } \leq \alpha$.
			\end{itemize}
		\end{definition}

		\begin{remark}\label{remark:dp-sanitizer-guarantees}
			Given an $(\epsilon, \delta, \alpha, \beta)$-differentially private sanitizer as in \cref{definition:dp-danitizer} one can replace the answer $\algo{B}{ \serverDS, \query }$ with $\textsc{B}^\prime ( \serverDS, \allowbreak \query ) = \algo{B}{ \serverDS, \query } + \alpha$.
			Hence, with probability $1 - \beta$, for all $\query \in \querySet$, $0 \leq \algo{\ensuremath{\textsc{B}^\prime}}{ \serverDS, \query } - \sum_i \query(d_i) \leq 2 \alpha$.
			We will hence assume from now on that sanitizers have this latter guarantee on their error.
		\end{remark}

		The main idea of \emph{sanitization} (a.k.a.\ private data release) is to release specific noisy statistics on a private dataset once, which can then be combined in order to answer an arbitrary number of queries without violating privacy.
		Depending on the query type and the notion of differential privacy (i.e., pure or approximate), different upper bounds on the error have been proven.
		Omitting the dependency on $\epsilon,\delta$, in case of point queries over domain size \domainSize{}, pure differential privacy results in $\alpha = \bigTheta{\log \domainSize}$~\cite{bounds-on-sample-complexity}, while for approximate differential privacy $\alpha = \bigO{1}$~\cite{private-learning-and-sanitization}.
		For range queries over domain size \domainSize{}, these bounds are $\alpha = \bigTheta{\log \domainSize}$ for pure differential privacy~\cite{non-interactive-database-privacy,dp-under-observation}, and $\alpha = \bigO{(\log^{*} \domainSize)^{1.5}}$ for approximate differential privacy (with an almost matching lower bound of $\alpha = \bigOmega{\log^{*} \domainSize}$)~\cite{private-learning-and-sanitization, dp-release, privately-learning-thresholds}.
		More generally, \citet{non-interactive-database-privacy} showed that any finite query set \querySet{} can be sanitized, albeit non-efficiently.

		% \log^{*} is "iterated logarithm" : https://en.wikipedia.org/wiki/iterated_logarithm

		\subsubsection*{Answering point and range queries with differential privacy}

			Utilizing the LPA for answering point queries results in error $\alpha = \bigO{\log \domainSize}$.
			A practical solution for answering range queries with error bounds very close to the optimal ones is the hierarchical method~\cite{dp-under-observation, accuracy-dp-histograms, dp-wavelet}.
			The main idea is to build an aggregate tree on the domain, and add noise to each node proportional to the tree height (i.e., noise scale logarithmic in the domain size \domainSize{}).
			Then, every range query is answered using the minimum number of tree nodes.
			\citet{hierarchical-methods-for-dp} showed that the hierarchical algorithm of \citet{accuracy-dp-histograms}, when combined with their proposed optimizations, offers the lowest error.

		\subsubsection*{Composition}

			Finally, we include a \emph{composition} theorem (adapted from \cite{privacy-integrated-queries}) based on \cite{differential-privacy-original,our-data-ourselves}. % chktex 2
			It concerns executions of multiple differentially private mechanisms on non-disjoint and disjoint inputs.

			\begin{theorem}\label{theorem:composition}
				Let \fromNtoM{\algo{A}}{1}{r} be mechanisms, such that each $\algo{A}_i$ provides $\epsilon_i$-differential privacy.
				Let \fromNtoM{\database}{1}{r} be pairwise non-disjoint (resp., disjoint) datasets.
				Let $\algo{A}$ be another mechanism that executes $\algo{A}_1(\database_1), \ldots, \algo{A}_r(\database_r)$ using independent randomness for each $\algo{A}_i$, and returns their outputs.
				Then, mechanism $\algo{A}$ is $\left( \sum_{i=1}^r \epsilon_i \right)$-differentially private (resp., $\left( \max_{i=1}^r \epsilon_i \right)$-differentially private).
			\end{theorem}

	\subsection{Oblivious RAM}\label{section:building-blocks:oram}

		Informally, Oblivious RAM (ORAM) is a mechanism that lets a user hide their RAM access pattern to remote storage.
		An adversarial server can monitor the actual accessed locations, but she cannot tell a read from a write, the content of the block or even whether the same logical location is being referenced.
		The notion was first defined by \citet{oram-theory} and \citet{oram-original}.

		More formally, a $(\eta_1, \eta_2)$-ORAM protocol is a two-party protocol between a user \user{} and a server \server{} who stores a RAM array.
		In each round, the user \user{} has input $(o, a, d)$, where $o$ is a RAM operation (\oramRead{} or \oramWrite{}), $a$ is a memory address and $d$ is a new data value, or $\bot$ for read operation.
		The input of \server{} is the current array.
		Via the protocol, the server updates the memory or returns to \user{} the data stored at the requested memory location, respectively.
		We speak of a sequence of such operations as a program \oramProgram{} being \emph{executed under the ORAM}.

		An ORAM protocol must satisfy correctness and security.
		Correctness requires that \user{} obtains the correct output of the computation except with at most probability $\eta_1$.
		For security, we require that for every user \user{} there exists a simulator $\simulator_\oram$ which provides a simulation of the server's view in the above experiment given only the number of operations.
		That is, the output distribution of $\simulator_\oram (c)$ is indistinguishable from $\algo{View}_\server$ with probability at most $\eta_2$ after $c$ protocol rounds.

		ORAM protocols are generally stateful, after each execution the client and server states are updated.
		\emph{For brevity, throughout the paper we will assume the ORAM state updates are implicit, including the encryption key \queryKey{} generated and maintained by the client.}

		Some existing efficient ORAM protocols are Square Root ORAM \cite{oram-theory}, Hierarchical ORAM \cite{oram-original}, Binary-Tree ORAM \cite{binary-tree-oram}, Interleave Buffer Shuffle Square Root ORAM \cite{shortest-path-oram}, TP-ORAM \cite{tp-oram}, Path-ORAM \cite{path-oram} and TaORAM \cite{taostore}. % chktex 2
		For detailed descriptions of each protocol, we recommend the work of \citet{oram-survey-feifei}.
		The latter three ORAMs achieve the lowest communication and storage overheads, $\bigO{\log \dataSize}$ and \bigO{\dataSize}, respectively.

	\section{Differentially private outsourced database systems}\label{section:dpodb}

	In this section we present our model, \emph{differentially private outsourced database system}, CDP-ODB, its security definition, query types and efficiency measures.
	It is an extension of the ODB model in \cref{section:background}.

	\subsection{Adversarial model}\label{section:dpodb:adversarial-models}

		We consider an honest-but-curious polynomial time adversary that attempts to breach differential privacy with respect to the input database \database{}.
		We observe later in \cref{section:dpodb:adversarial-models:adaptive} that it is impossible to completely hide the number of records returned on each query without essentially returning all the database records on each query.
		This, in turn, means that different query sequences may be distinguished, and, furthermore, that differential privacy may not be preserved if the query sequence depends on the content of the database records.
		We hence, only require the protection of differential privacy with respect to every fixed query sequence.
		Furthermore, we relax to computational differential privacy (following~\cite{computational-dp}).

		In the following definition, the notation \view{\protocol{}}{\database, \fromNtoM{\query}{1}{m}} denotes the view of the server \server{} in the execution of protocol \protocol{} in answering queries \fromNtoM{\query}{1}{m} with the underlying database \database{}.

		\begin{definition}
			We say that an outsourced database system \protocol{} is $(\epsilon, \delta)$-computationally differentially private (a.k.a.~CDP-ODB) if for every polynomial time distinguishing adversary \adversary{}, for every neighboring databases $\database \sim \database^\prime$, and for every query sequence $\fromNtoM{\query}{1}{m} \in \querySet^m$ where $m = \mathsf{poly}(\lambda)$,

			\begin{multline*}
				\probability{\adversary \left( 1^\lambda, \view{\protocol{}}{\database, \fromNtoM{\query}{1}{m}} \right) = 1 } \leq \\
				\exp{\epsilon} \cdot \probability{\adversary \left( 1^\lambda, \view{\protocol{}}{\database^\prime, \fromNtoM{\query}{1}{m}} \right) = 1} + \delta +\negl \; ,
			\end{multline*}
			where the probability is over the randomness of the distinguishing adversary \adversary{} and the protocol \protocol{}.
		\end{definition}

	\begin{remark}[Informal]
		We note that security and differential privacy in this model imply protection against communication volume and access pattern leakages and thus prevent a range of attacks, such as \cite{leakage-abuse-attacks-cash-15,inference-attacks-naveed-15,generic-attacks-kellaris}. % chktex 2
	\end{remark}

	\subsubsection{On impossibility of adaptive queries}\label{section:dpodb:adversarial-models:adaptive}

		Non-adaptivity in our CDP-ODB definition does not reflect a deficiency of our specific protocol but rather an inherent source of leakage when the queries may depend on the decrypted data.
		Consider an adaptive CDP-ODB definition that does not fix the query sequence \fromNtoM{q}{1}{m} in advance but instead an arbitrary (efficient) user \user{} chooses them during the protocol execution with \server{}.
		As before, we ask that the \server{}'s view is DP on neighboring databases for every such \user{}.
		We observe that this definition cannot possibly be satisfied by \emph{any} outsourced database system without unacceptable efficiency overhead.
		Note that non-adaptivity here does not imply that the client knows all the queries in advance, but rather can choose them at any time (e.g., depending on external circumstances) as long as they do not depend on true answers to prior queries.

		To see this, consider two neighboring databases $\database, \database^\prime$.
		Database \database{} has 1 record with $\mathsf{key} = 0$ and $\database^\prime$ has none.
		Furthermore, both have 50 records with $\mathsf{key} = 50$ and 100 records with $\mathsf{key} = 100$.
		User \user{} queries first for the records with $\mathsf{key} = 0$, and then if there is a record with $\mathsf{key} = 0$ it queries for the records with $\mathsf{key} = 50$, otherwise for the records with $\mathsf{key} = 100$.
		Clearly, an efficient outsourced database system cannot return nearly as many records when $\mathsf{key} = 50$ versus $\mathsf{key} = 100$ here.
		Hence, this allows distinguishing $\database, \database^\prime$ with probability almost 1.

		To give a concrete scenario, suppose neighboring medical databases differ in one record with a rare diagnosis ``Alzheimer's disease''.
		A medical professional queries the database for that diagnosis first (point query), and if there is a record, she queries the senior patients next (range query, \texttt{age $\ge$ 65}), otherwise she queries the general population (resulting in more records).
		We leave it open to meaningfully strengthen our definition while avoiding such impossibility results, and we defer the formal proof to future work.

	% chktex-file 1
% chktex-file 8
% chktex-file 21
% chktex-file 24
% chktex-file 26
% chktex-file 36
% chktex-file 37

\newlength{\setupLength}
\setlength{\setupLength}{5.5em}
\newlength{\queryLength}
\setlength{\queryLength}{9.5em}

\begin{algorithm*}[t]

	\begin{pchstack}

		\procedure[linenumbering]{\protocolSetup{}}{
													\textbf{User \user}																\>																\> \textbf{Server \server}	\\
			\label{algorithm:dp-oram:setup:line-2}	\pcinput{\database}																\>																\> \pcinput{\emptyset}		\\
			\label{algorithm:dp-oram:setup:line-3}	\indexI \gets \algo{CreateIndex}{\database}										\>																\>							\\
			\label{algorithm:dp-oram:setup:line-4}	\oramProgram = \left. (\oramWrite, \recordID_i, \record_i) \right|_{i = 1}^n	\>																\>							\\
			\label{algorithm:dp-oram:setup:line-5}																					\> \sendmessageboth*[\setupLength]{\algo{ORAM}{\oramProgram}}	\>							\\
			\label{algorithm:dp-oram:setup:line-6}	\serverDS \gets \algo{A}{\fromNtoM{\searchKey}{1}{\domainSize}}					\> \sendmessageright*[\setupLength]{\serverDS}					\>							\\
			\label{algorithm:dp-oram:setup:line-7}	\pcouput{\indexI}																\>																\> \pcouput{\serverDS}
		}

		\hspace{2em}

		\procedure[linenumbering]{\protocolQuery{}}{
													\textbf{User \user}																				\>																											\> \textbf{Server \server}				\\
			\label{algorithm:dp-oram:query:line-2}	\pcinput{\query, \indexI}																		\>																											\> \pcinput{\serverDS}					\\
			\label{algorithm:dp-oram:query:line-3}	T \gets \algo{Lookup}{\indexI, \query}															\> \sendmessageright*[\queryLength]{\query}																	\> c \gets \algo{B}{\serverDS, \query}	\\
			\label{algorithm:dp-oram:query:line-4}	\oramProgram_\mathsf{true} = \left. (\oramRead, \recordID_i, \bot) \right|_{i \in T}			\> \sendmessageleft*[\queryLength]{c}																		\>										\\
			\label{algorithm:dp-oram:query:line-5}	\oramProgram_\mathsf{noise} = \left. (\oramRead, S \setminus T, \bot) \right|_{1}^{c - \abs{T}}	\>																											\>										\\
			\label{algorithm:dp-oram:query:line-6}	R																								\> \sendmessageboth*[\queryLength]{\algo{ORAM}{\oramProgram_\mathsf{true} \| \oramProgram_\mathsf{noise}}}	\>										\\
			\label{algorithm:dp-oram:query:line-7}	\pcouput{R}																						\>																											\>	\pcouput{\emptyset}
		}

	\end{pchstack}

	\caption{
		\epsolute{} protocol.
		$\algo{ORAM}{\cdot}$ denotes an execution of ORAM protocol (\cref{section:building-blocks:oram}), where \user{} plays the role of the client.
		ORAM protocol client and server states are implicit.
		$S \setminus T$ represents a set of valid record IDs $S$ that are not in the true result set $T$.
	}%
	\label{algorithm:dp-oram}
\end{algorithm*}

	\subsection{Query types}

		In this work we are concerned with the following query types:
		\begin{description}[style=unboxed, leftmargin=0em]
			\item[Range queries]
				Here we assume a total ordering on \searchKeyDomain{}.
				A query \query{a}{b} is associated with an interval $\interval{a}{b}$ for $1 \leq a \leq b \leq \domainSize$ such that $\query{a}{b}(c) = 1$ iff $c \in \interval{a}{b}$ for all $c \in \searchKeyDomain$.
				The equivalent SQL query is:

				\smallskip
				\indent\texttt{SELECT * FROM table WHERE attribute BETWEEN a AND b;}
				\smallskip

			\item[Point queries]
				Here \searchKeyDomain{} is arbitrary and a query predicate \query{a} is associated with an element $a \in \searchKeyDomain$ such that $\query{a}(b) = 1$ iff $a = b$.
				In an ordered domain, point queries are degenerate range queries.
				The equivalent SQL query is:

				\smallskip
				\indent\texttt{SELECT * FROM table WHERE attribute = a;}

		\end{description}

	\subsection{Measuring Efficiency}

		We define two basic efficiency measures for a CDP-ODB\@.
		\begin{description}[style=unboxed, leftmargin=0em]
			\item[Storage efficiency]
				is defined as the sum of the bit-lengths of the records in a database relative to the bit-length of a corresponding encrypted database.
				Specifically, we say that an outsourced database system has \emph{storage efficiency} of $(\efficiencyCoefficient, \efficiencyOffset)$ if the following holds.
				Fix any \databaseDef{} and let $n_1 = \sum_{i=1}^n \abs{r_i}$.
				Let $\server_\mathsf{state}$ be an output of \server{} on a run of \protocolSetup{} where \user{} has input \database{}, and let $n_2 = \abs{ \server_\mathsf{state} }$.
				Then $n_2 \leq \efficiencyCoefficient n_1 + \efficiencyOffset$.

			\item[Communication efficiency]
				is defined as the sum of the lengths of the records in bits whose search keys satisfy the query relative to the actual number of bits sent back as the result of a query.
				Specifically, we say that an outsourced database system has \emph{communication efficiency} of $(\efficiencyCoefficient, \efficiencyOffset)$ if the following holds.
				Fix any \query{} and \serverDS{} output by \protocolSetup{}, let \user{} and \server{} execute \protocolQuery{} where \user{} has inputs \query{}, and output $R$, and \server{} has input \serverDS{}.
				Let $m_1$ be the amount of data in bits transferred between \user{} and \server{} during the execution of \protocolQuery{}, and let $m_2 = \abs{ R }$.
				Then $m_2 \leq \efficiencyCoefficient m_1 + \efficiencyOffset$.
		\end{description}

		Note that $\efficiencyCoefficient \geq 1$ and $\efficiencyOffset \geq 0$ for both measures.
		We say that an outsourced database system is \emph{optimally storage efficient} (resp., \emph{optimally communication efficient}) if it has storage (resp., communication) efficiency of $(1, 0)$.

	\section{\texorpdfstring{\epsolute{}}{Epsolute}}\label{section:dp-oram}

	In this section we present a construction, \epsolute{}, that satisfies the security definition in \cref{section:dpodb}, detailing algorithms for both range and point query types.
	We also provide efficiency guarantees for approximate and pure DP versions of \epsolute{}.

	\subsection{General construction}

		Let \querySet{} be a collection of queries.
		We are interested in building a differentially private outsourced database system for \querySet{}, called \epsolute{}.
		Our solution will use these building blocks.
		\begin{itemize}
			\item
				A $(\eta_1, \eta_2)$-ORAM protocol \algo{ORAM}{\cdot}.
			\item
				An $(\epsilon, \delta, \alpha, \beta)$-differentially private sanitizer $(\algo{A}, \algo{B})$ for \querySet{} and negligible $\beta$, which satisfies the non-negative noise guarantee from \cref{remark:dp-sanitizer-guarantees}.
			\item
				A pair of algorithms \algo{CreateIndex} and \algo{Lookup}.
				\algo{CreateIndex} consumes \database{} and produces an index data structure \indexI{} that maps a search key \searchKey{} to a list of record IDs \recordID{} corresponding to the given search key.
				\algo{Lookup} consumes \indexI{} and \query{} and returns a list $T = \fromNtoM{\recordID}{1}{{\abs{T}}}$ of record IDs matching the supplied query.
		\end{itemize}

\begin{figure}[!ht]
	\centering
	\includegraphics[width=\linewidth]{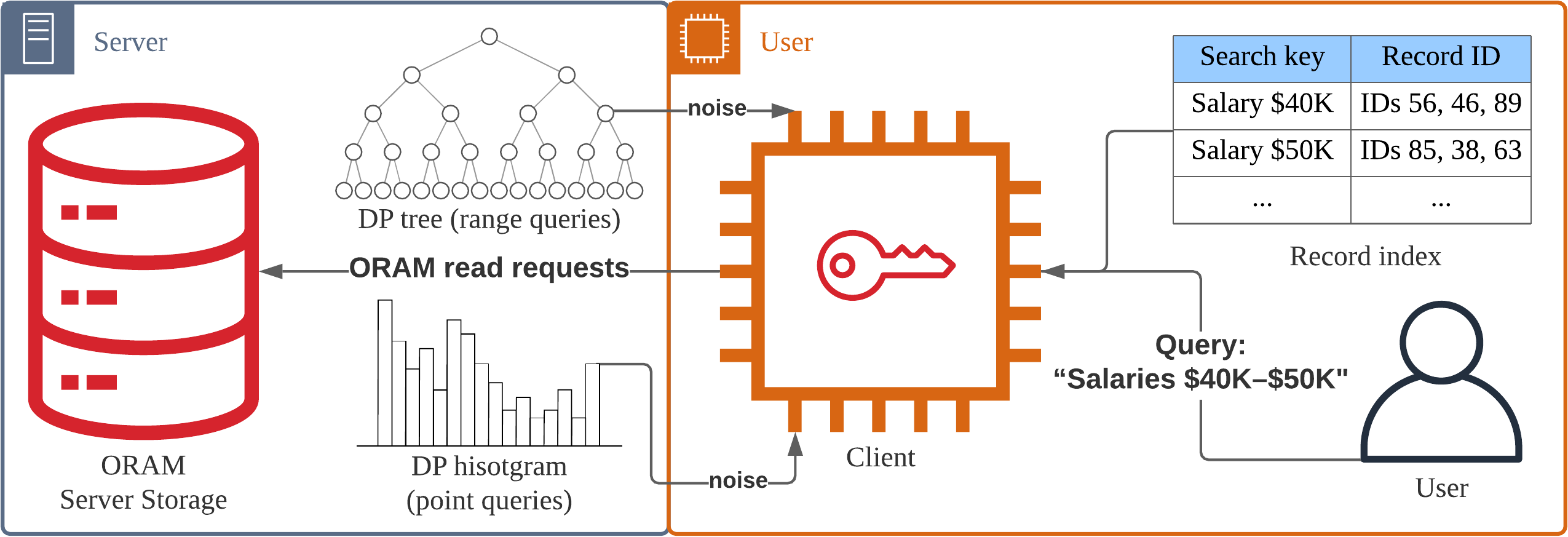}
	\Description{\epsolute{} construction}
	\caption{\epsolute{} construction}%
	\label{figure:dp-oram}
\end{figure}

		Our protocol $\protocol = (\protocolSetup, \protocolQuery)$ of \epsolute{} works as shown in \cref{algorithm:dp-oram}.
		Hereafter, we reference lines in \cref{algorithm:dp-oram}.
		See \cref{figure:dp-oram} for a schematic description of the protocol.

		\paragraph*{Setup protocol \; \texorpdfstring{\protocolSetup{}}{}}

			Let \user{}'s input be a database \databaseDef{} (\cref{algorithm:dp-oram:setup:line-2}).
			\user{} creates an index \indexI{} mapping search keys to record IDs corresponding to these keys (\cref{algorithm:dp-oram:setup:line-3}).
			\user{} sends over the records to \server{} by executing the ORAM protocol on the specified sequence (\crefrange{algorithm:dp-oram:setup:line-4}{algorithm:dp-oram:setup:line-5}).
			\user{} generates a DP structure \serverDS{} over the search keys using sanitizer \algo{A}, and sends \serverDS{} over to \server{} (\cref{algorithm:dp-oram:setup:line-6}).
			The output of \user{} is \indexI{} and of \server{} is \serverDS{}; final \algo{ORAM} states of \server{} and \user{} are implicit, including encryption key \queryKey{} (\cref{algorithm:dp-oram:setup:line-7}).

		\paragraph*{Query protocol \; \texorpdfstring{\protocolQuery{}}{}}

			\user{} starts with a query \query{} and index \indexI{}, \server{} starts with a DP structure \serverDS{}.
			One can think of these inputs as outputs of \protocolSetup{} (\cref{algorithm:dp-oram:query:line-2}).
			\user{} immediately sends the query to \server{}, which uses the sanitizer \algo{B} to compute the total number of requests $c$, while \user{} uses index \indexI{} to derive the true indices of the records the query \query{} targets (\cref{algorithm:dp-oram:query:line-3}).
			\user{} receives $c$ from \server{} and prepares two ORAM sequences: $\oramProgram_\mathsf{true}$ for real records retrieval, and $\oramProgram_\mathsf{noise}$ to pad the number of requests to $c$ to perturb the communication volume.
			$\oramProgram_\mathsf{noise}$ includes valid non-repeating record IDs that are not part of the true result set $T$ (\crefrange{algorithm:dp-oram:query:line-4}{algorithm:dp-oram:query:line-5}).
			\user{} fetches the records, both real and fake, from \server{} using the ORAM protocol (\cref{algorithm:dp-oram:query:line-6}).
			The output of \user{} is the filtered set of records requested by the query $\query{}$; final \algo{ORAM} states of \server{} and \user{} are implicit (\cref{algorithm:dp-oram:query:line-7}).

		The protocols for point and range queries only differ in sanitizer implementations, see \cref{section:dp-oram:point,section:dp-oram:range}.
		Note above that in any execution of \protocolQuery{} we have $c \geq \query(\database)$ with overwhelming probability $1 - \beta$ (by using sanitizers satisfying \cref{remark:dp-sanitizer-guarantees}), and thus the protocol is well-defined and its accuracy is $1 - \beta$.
		Also note that the DP parameter $\delta$ is lower-bounded by $\beta$ because sampling negative noise, however improbable, violates privacy, and therefore the final construction is $(\epsilon, \beta)$-DP\@.

	\subsection{Security}

		\begin{theorem}
			\epsolute{} is $(\beta \cdot m)$-wrong and $(\epsilon, \delta)$-CDP-ODB where the negligible term is $\negl = 2 \cdot \eta_2$.
		\end{theorem}

		\begin{proof}
			We consider a sequence of views
			\[
				\view{1} \to \view{2} \to \view{3} \to \view{4} \; .
			\]
			\view{1} is \view{\protocol}{\database, \fromNtoM{\query}{1}{m}}.
			\view{2} is produced only from $\serverDS \gets \algo{A}{\fromNtoM{\searchKey}{1}{\domainSize}}$.
			Namely, compute $c_i \gets \algo{A}{\serverDS, \query_i}$ for all $i$ and run ORAM simulator on $\sum_i c_i$.
			By ORAM security,
			\[
				\probability{\adversary(\view{1})} - \probability{\adversary(\view{2})} \leq \eta_2 \; .
			\]
			\view{3} is produced similarly but $\serverDS \gets \algo{A}{\fromNtoM{\searchKey^\prime}{1}{\domainSize}}$ instead.
			Note that the $c_i$ are simply post-processing on \serverDS{} via \algo{B} so
			\[
				\probability{\adversary(\view{2})} = \exp(\epsilon) \cdot \probability{\adversary(\view{3})} + \delta \; .
			\]
			$\view{4} = \view{\protocol}{\database^\prime, \fromNtoM{\query}{1}{m}}$.
			It follows by ORAM security
			\[
				\probability{\adversary(\view{3})} - \probability{\adversary(\view{4})} \leq \eta_2 \; .
			\]
			Putting this all together completes the proof.
		\end{proof}

	\subsection{Efficiency}

		For an ORAM with communication efficiency $(a_1, a_2)$ and an $(\alpha, \beta)$-differentially private sanitizer, the \epsolute{} communication efficiency is $(a_1, a_2 \cdot \alpha)$.
		The efficiency metrics demonstrate how the total storage or communication volume (the number of stored or transferred bits) changes additively and multiplicatively as the functions of data size \dataSize{} and domain \domainSize{}.
		We therefore have the following corollaries for the efficiency of the system in the cases of approximate and pure differential privacy.
		\begin{corollary}\label{corollary:comm-efficiency-approximate-dp}
			\epsolute{} is an outsourced database system with storage efficiency \efficiency{1}{0}.
			Depending on the query type, assume it offers the following communication efficiency.
			\begin{description}
				\item[Range queries] $\efficiency{\log \dataSize}{2^{\log^* \domainSize} \log \dataSize}$
				\item[Point queries] $\efficiency{\log \dataSize}{\log \dataSize}$
			\end{description}
			Then, there is a negligible $\delta$ such that \epsolute{} satisfies $(\epsilon, \delta)$\hyp{}differential privacy for some $\epsilon$.\footnote{
				Note that the existence of $\epsilon$ in this setting implies that the probability of an adversary breaking the DP guarantees is bounded by it.
			}
		\end{corollary}

		\begin{proof}
			By using ORAM, we store only the original data once and hence, we get optimal storage efficiency.

			The communication efficiency depends on the upper bound of the error for each sanitizer when $\delta > 0$, as described in \cref{section:building-blocks:dp} and \cref{remark:dp-sanitizer-guarantees}.
			The most efficient ORAM protocol to date has $\bigO{\log \dataSize}$ communication overhead (see \cref{section:building-blocks:oram}).
		\end{proof}

		\begin{corollary}\label{corollary:comm-efficiency-pure-dp}
			\epsolute{} is an outsourced database system with storage efficiency \efficiency{1}{0}.
			Depending on the query type, assume it offers the following communication efficiency.
			\begin{description}
				\item[Range queries] $\efficiency{\log \dataSize}{\log \domainSize \log \dataSize}$
				\item[Point queries] $\efficiency{\log \dataSize}{\log \domainSize \log \dataSize}$
			\end{description}
			Then, \epsolute{} satisfies $\epsilon$-differential privacy for some $\epsilon$.
		\end{corollary}

		\begin{proof}
			Similarly, we derive the proof by considering the use of ORAM and the upper bound of the error for each sanitizer when $\delta = 0$ in \cref{section:building-blocks:dp}.
		\end{proof}

	\subsection{Extending to multiple attributes}\label{section:dp-oram:multiple-attributes}

		We will now describe how \epsolute{} supports multiple indexed attributes and what the privacy and performance implications are.
		The na\"{\i}ve way is to simply duplicate the entire stack of states of \user{} and \server{}, and during the query use the states whose attribute the query targets.
		However, \epsolute{} design allows to keep the most expensive part of the state --- the ORAM state --- shared for all attributes and both types of queries.
		Specifically, the index \indexI{} and DP structure \serverDS{} are generated per attribute and query type, while \user{} and \server{} ORAM states are generated once.
		This design is practical since \serverDS{} is tiny and index \indexI{} is relatively small compared to ORAM states, see \cref{section:experiments}.

		We note that in case the indices grow large in number, it is practical to outsource them to the adversarial server using ORAM and download only the ones needed for each query.
		In terms of privacy, the solution is equivalent to operating different \epsolute{} instances because ORAM hides the values of records and access patterns entirely.
		Due to \cref{theorem:composition} for non-disjoint datasets, the total privacy budget of the multi-attribute system will be the sum of individual budgets for each attribute / index.

		Next, we choose two DP sanitizers for our system, for point and for range queries, and calculate the $\alpha$ values to make them output positive values with high probability, consistent with \cref{remark:dp-sanitizer-guarantees}.

	\subsection{\texorpdfstring{\epsolute{}}{Epsolute} for point queries}\label{section:dp-oram:point}

		For point queries, we use the LPA method as the sanitizer to ensure pure differential privacy.
		Specifically, for every histogram bin, we draw noise from the Laplace distribution with mean $\alpha_p$ and scale $\lambda = \nicefrac{1}{\epsilon}$.
		To satisfy \cref{remark:dp-sanitizer-guarantees}, we have to set $\alpha_p$ such that if values are drawn from $\algo{Laplace}{ \alpha_p, \nicefrac{1}{\epsilon} }$ at least as many times as the number of bins \domainSize{}, they are all positive with high probability $1 - \beta$, for negligible $\beta$.

		We can compute the exact minimum required value of $\alpha_p$ in order to ensure drawing positive values with high probability by using the CDF of the Laplace distribution.
		Specifically, $\alpha_p$ should be equal to the minimum value that satisfies the following inequality.

		\[
			\left( 1 - \frac{1}{2} e^{- \alpha_p \cdot \epsilon} \right)^\domainSize \leq 1 - \beta
		\]
		which is equivalent to
		\[
			\alpha_p = \ceil{ -\frac{ \ln \left( 2 - 2 \sqrt[\domainSize]{1 - \beta} \right) }{ \epsilon } }
		\]

	% chktex-file 1
% chktex-file 8
% chktex-file 21
% chktex-file 24
% chktex-file 26
% chktex-file 36
% chktex-file 37

\setlength{\setupLength}{5em}
\setlength{\queryLength}{9em}

\newcommand{\SetupGamma}{
	\procedure[linenumbering]{\protocolSetup{} of \protocolGamma{}}{
																\textbf{User \user}																							\>																\> \textbf{Server \server}	\\
		\label{algorithm:dp-oram-parallel:gamma:setup:line-2}	\pcinput{\database{}}																						\>																\> \pcinput{\emptyset}		\\
		\label{algorithm:dp-oram-parallel:gamma:setup:line-3}	\indexI \gets \algo{CreateIndex}{\database, \oramsNumber}													\>																\>							\pclb
		\pcintertext[dotted]{$\pcfor j \in \set{1, \ldots, \oramsNumber} \pcdo$ \; \text{(in parallel)}}
		\label{algorithm:dp-oram-parallel:gamma:setup:line-4}	\left\langle \overline{\record}, \overline{\recordID} \right\rangle\ \text{s.t.}\ \algo{H}{\recordID} = j	\>																\>							\\
		\label{algorithm:dp-oram-parallel:gamma:setup:line-5}	\oramProgram = \left\langle (\oramWrite, \overline{\recordID}, \overline{\record}) \right\rangle			\>																\>							\\
		\label{algorithm:dp-oram-parallel:gamma:setup:line-6}																												\> \sendmessageboth*[\setupLength]{\algo{ORAM}_j(\oramProgram)}	\>							\pclb
		\pcintertext[dotted]{$\pcendfor$}
		\label{algorithm:dp-oram-parallel:gamma:setup:line-7}	\serverDS \gets \algo{A}{\fromNtoM{\searchKey}{1}{\domainSize}}												\> \sendmessageright*[\setupLength]{\serverDS}					\>							\\
		\label{algorithm:dp-oram-parallel:gamma:setup:line-8}	\pcouput{\indexI}																							\>																\> \pcouput{ \serverDS }
	}
}

\newcommand{\QueryGamma}{
	\procedure[linenumbering]{\protocolQuery{} of \protocolGamma{}}{
																\textbf{User \user}																					\>																												\> \textbf{Server \server}									\\
		\label{algorithm:dp-oram-parallel:gamma:query:line-2}	\pcinput{\query, \indexI}																			\>																												\> \pcinput{ \serverDS }									\\
		\label{algorithm:dp-oram-parallel:gamma:query:line-3}	\fromNtoM{T}{1}{\oramsNumber} \gets \algo{Lookup}{I, \query}										\> \sendmessageright*[\queryLength]{\query}																		\> k \gets \algo{B}{\serverDS, \query}						\\
		\label{algorithm:dp-oram-parallel:gamma:query:line-4}																										\> \sendmessageleft*[\queryLength]{c}																			\> c \gets (1 + \gamma) \frac{\tilde{k}_0}{\oramsNumber}	\pclb
		\pcintertext[dotted]{$\pcfor j \in \set{1, \ldots, \oramsNumber} \pcdo$ \; \text{(in parallel)}}
		\label{algorithm:dp-oram-parallel:gamma:query:line-5}	\oramProgram_\mathsf{true} = \left. (\oramRead, \recordID_i, \bot) \right|_{i \in T_j}				\>																												\>															\\
		\label{algorithm:dp-oram-parallel:gamma:query:line-6}	\oramProgram_\mathsf{noise} = \left. (\oramRead, S \setminus T_j, \bot) \right|_{1}^{c - \abs{T_j}}	\>																												\>															\\
		\label{algorithm:dp-oram-parallel:gamma:query:line-7}	R_j																									\> \sendmessageboth*[\queryLength]{\algo{ORAM}_j(\oramProgram_\mathsf{true} \| \oramProgram_\mathsf{noise})}	\>															\pclb
		\pcintertext[dotted]{$\pcendfor$}
		\label{algorithm:dp-oram-parallel:gamma:query:line-8}	\pcouput{ \left. R_j \right|_{j = 1}^\oramsNumber }													\>																												\>	\pcouput{\emptyset}
	}
}

\begin{algorithm*}[t]

	\begin{pcvstack}

		\begin{pchstack}

			\SetupGamma{}

			\hspace{5pt}

			\QueryGamma{}

		\end{pchstack}

	\end{pcvstack}

	\caption{
		Parallel \epsolute{} for \protocolGamma{}, extends \cref{algorithm:dp-oram}.
		\oramsNumber{} is the number of parallel ORAMs.
		\algo{H} is a random hash function $\algo{H} : \bin^* \to \set{1, \ldots, \oramsNumber}$.
		$\gamma$ and $\tilde{k}_0$ are computed as in \cref{section:prallel-dp-oram:gamma}.
		\user{} and \server{} maintain \oramsNumber{} ORAM states implicitly.
	}%
	\label{algorithm:dp-oram-parallel}
\end{algorithm*}

	\subsection{\texorpdfstring{\epsolute{}}{Epsolute} for range queries}\label{section:dp-oram:range}

		For range queries, we implement the aggregate tree method as the sanitizer.
		Specifically, we build a complete \fanout{}-ary tree on the domain, for a given \fanout{}.
		A leaf node holds the number of records falling into each bin plus some noise.
		A parent node holds sum of the leaf values in the range covered by this node, plus noise.
		Every time a query is issued, we find the minimum number of nodes that cover the range, and determine the required number of returned records by summing these node values.
		Then, we ask the server to retrieve the records in the range, plus to retrieve multiple random records so that the total number of retrieved records matches the required number of returned records.

		The noise per node is drawn from the Laplace distribution with mean $\alpha_h$ and scale $\lambda = \frac{\log_{\fanout} \domainSize}{\epsilon}$.
		Consistent with \cref{remark:dp-sanitizer-guarantees}, we determine the mean value $\alpha_h$ in order to avoid drawing negative values with high probability.
		We have to set $\alpha_h$ such that if values are drawn from $\algo{Laplace}{ \alpha_h, \frac{\log_{\fanout} \domainSize}{\epsilon} }$ at least as many times as the number of nodes in the tree, they are all positive with high probability $1 - \beta$, for negligible $\beta$.

		Again, we can compute the exact minimum required value of $\alpha_h$ in order to ensure drawing positive values with high probability by using the CDF of the Laplace distribution.
		Specifically, $\alpha_h$ should be equal to the minimum value that satisfies the following inequality.
		\[
			\left( 1 - \frac{1}{2} e^{- \frac{\alpha_h \cdot \epsilon}{\log_{\fanout} \domainSize}} \right)^\mathsf{nodes} \leq 1 - \beta
		\]
		which is equivalent to
		\begin{equation}\label{equation:min-mu-for-range}
			\alpha_h = \ceil{ -\frac{ \ln{ (2 - 2 \sqrt[\mathsf{nodes}]{ 1 - \beta } ) } \cdot \log_{\fanout} \domainSize }{\epsilon} }
		\end{equation}
		where $\mathsf{nodes} = \frac{ \fanout^{ \ceil{ \log_{\fanout} (\fanout - 1) + \log_{\fanout} \domainSize - 1 }} - 1}{ \fanout - 1 } + \domainSize$ is the total number of tree nodes.

	\section{An efficient Parallel \texorpdfstring{\epsolute{}}{Epsolute}}\label{section:prallel-dp-oram}

	While the previously described scheme is a secure and correct CDP-ODB, a single-threaded implementation may be prohibitively slow in practice.
	To bring the performance closer to real-world requirements, we need to be able to scale the algorithm horizontally.
	In this section, we describe an upgrade of \epsolute{} --- a scalable parallel solution.

	We suggest two variants of parallel \epsolute{} protocol.
	Both of them work by operating \oramsNumber{} ORAMs and randomly assigning to each of them $\nicefrac{n}{\oramsNumber}$ database records.
	For each query, we utilize the index \indexI{} to find the required records from the corresponding ORAMs.
	For each ORAM, we execute a separate thread to retrieve the records.
	The threads work in parallel and there is no need for locking, since each ORAM works independently from the rest.
	We present two methods that differ in the way they build and store DP structure \serverDS{}, and hence the number of ORAM requests they make.

	\subsection{\texorpdfstring{No-$\gamma$-method}{No-gamma-method}: DP structure per ORAM}

		In \protocolNoGamma{}, for each ORAM / subset of the dataset, we build a DP index the same way as described in \cref{section:dp-oram}.
		We note that \cref{theorem:composition} for disjoint datasets applies to this construction: the privacy budget $\epsilon$ for the construction is the largest (least private) among the $\epsilon$'s of the DP indices for each ORAM / subset of the dataset.

		The communication efficiency changes because
		\begin{enumerate*}[label={(\roman*)}]
			\item
				we essentially add \oramsNumber{} record subsets in order to answer a query, each having at most $\alpha$ extra random records, and
			\item
				each ORAM holds fewer records than before, resulting in a tree of height $\log \frac{\dataSize}{\oramsNumber}$.
		\end{enumerate*}

		However, we cannot expect that the records required for each query are equally distributed among the different ORAMs in order to reduce the multiplicative communication cost from $\log \dataSize$ to $\frac{\log \dataSize}{\oramsNumber}$.
		Instead, we need to bound the worst case scenario which is represented by the maximum number of records from any ORAM that is required to answer a query.
		This can be computed as follows.

		Let $X_j$ be $1$ if a record for answering query \query{} is in a specific $\algo{ORAM}_j$, and $0$ otherwise.
		Due to the random assignment of records to ORAMs, $\probability{X_j = 1} = \nicefrac{1}{\oramsNumber}$.
		Assume that we need $k_0$ records in order to answer query \query{}.
		The maximum number of records from $\algo{ORAM}_j$ in order to answer \query{} is bounded as follows.

		\begin{equation}\label{equation:gamma}
			\probability{ \sum_{i=1}^{k_0} X_i > ( 1 + \gamma ) \frac{k_0}{\oramsNumber} } \leq \exp{ \left( - \frac{ k_0 \gamma^2 }{ 3 \oramsNumber } \right) }
		\end{equation}

		Finally, we need to determine the value of $\gamma$ such that $\exp{ \left( - \frac{ k_0 \gamma^2 }{ 3 \oramsNumber } \right) }$ is smaller than the value $\beta$.
		Thus, $\gamma = \sqrt{ \frac{-3 \oramsNumber \log \beta}{ k_0 } }$.
		The communication efficiency for each query type is described in the following corollary.

		\begin{corollary}\label{corollary:no-gamma}
			Let \protocolNoGamma{} be an outsourced database system with storage efficiency \efficiency{1}{0}.
			Depending on the query type, \protocolNoGamma{} offers the following communication efficiency.
			\begin{description}
				\item[Range queries] $\efficiency{\left( 1 + \sqrt{ \frac{- 3 \oramsNumber \log \beta}{k_0} } \right) \log \frac{\dataSize}{\oramsNumber} }{ \frac{ \log^{1.5} \domainSize}{ \epsilon } \oramsNumber \log \dataSize }$
				\item[Point queries] $\efficiency{\left( 1 + \sqrt{ \frac{- 3 \oramsNumber \log \beta}{k_0} } \right) \log \frac{\dataSize}{\oramsNumber} }{ \frac{ \log \domainSize}{ \epsilon } \oramsNumber \log \dataSize }$
			\end{description}

			Then, \protocolNoGamma{} satisfies $\epsilon$-differential privacy for some $\epsilon$.
		\end{corollary}

		In our experiments, we set \oramsNumber{} as a constant depending on the infrastructure.
		However, if \oramsNumber{} is set as $\bigO{\log n}$, the total communication overhead of the construction will still exceed the lower-bound presented in~\cite{multi-server-orams}.

	\subsection{\texorpdfstring{$\gamma$-method}{Gamma-method}: shared DP structure}\label{section:prallel-dp-oram:gamma}

		In \protocolGamma{}, we maintain a single shared DP structure \serverDS{}.
		When a query is issued, we must ensure that the number of records retrieved from every ORAM is the same.
		As such, depending on the required noisy number of records $\tilde{k}_0$, we need to retrieve at most $( 1 + \gamma ) \frac{ \tilde{k}_0 }{\oramsNumber}$ records from each ORAM, see \cref{equation:gamma}, for $\gamma = \sqrt{ \frac{-3 \oramsNumber \log \beta}{ \tilde{k}_0 } }$.
		Setting $\tilde{k}_0 = k_0 + \frac{\log^{1.5} \domainSize}{\epsilon}$ for range queries and $\tilde{k}_0 = k_0 + \frac{\log \domainSize}{\epsilon}$ for point queries, the communication efficiency is as follows.

		\begin{corollary}\label{corollary:gamma}
			Let \protocolGamma{} be an outsourced database system with storage efficiency \efficiency{1}{0}.
			Depending on the query type, \protocolGamma{} offers the following communication efficiency.
			\begin{description}
				\item[Range queries] $\efficiency{ \left( 1 + \sqrt{ \frac{-3 \oramsNumber \log \beta}{k_0 + \frac{ \log^{1.5} \domainSize }{ \epsilon }} }\right) \log \frac{\dataSize}{\oramsNumber} \left( 1 + \frac{\log^{1.5} \domainSize}{\epsilon} \right) }{0}$
				\item[Point queries] $\efficiency{ \left( 1 + \sqrt{ \frac{-3 \oramsNumber \log \beta}{k_0 + \frac{ \log \domainSize }{ \epsilon }} }\right) \log \frac{\dataSize}{\oramsNumber} \left( 1 + \frac{\log \domainSize}{\epsilon} \right) }{0}$
			\end{description}
			Then, \protocolGamma{} satisfies $\epsilon$-differential privacy for some $\epsilon$.
		\end{corollary}

		\protocolGamma{} is depicted in \cref{algorithm:dp-oram-parallel}.
		There are a few extensions to the subroutines and notation from \cref{algorithm:dp-oram}.
		\algo{CreateIndex} and \algo{Lookup} now build and query the index which maps a search key to a pair --- the record ID and the ORAM ID (1 to \oramsNumber{}) which stores the record.
		\Crefrange{algorithm:dp-oram-parallel:gamma:setup:line-4}{algorithm:dp-oram-parallel:gamma:setup:line-6} of \cref{algorithm:dp-oram-parallel} \protocolSetup{} repeat for each ORAM and operate on the records partitioned for the given ORAM using hash function \algo{H} on the record ID\@.
		A shared DP structure is created with the sanitizer \algo{A} (\cref{algorithm:dp-oram-parallel:gamma:setup:line-7}).
		In \cref{algorithm:dp-oram-parallel} \protocolQuery{}, the total number of ORAM requests is computed once (\cref{algorithm:dp-oram-parallel:gamma:query:line-4}).
		\Crefrange{algorithm:dp-oram-parallel:gamma:query:line-5}{algorithm:dp-oram-parallel:gamma:query:line-7} repeat for each ORAM and operate on the subset of records stored in the given ORAM\@.
		Note that \user{} and \server{} implicitly maintain \oramsNumber{} ORAM states, and the algorithm uses the $(\algo{A}, \algo{B})$ sanitizer defined in \cref{section:dp-oram}.

		Note that we guarantee privacy and access pattern protection on a record level.
		Each ORAM gets accessed at least once (much more than once for a typical query) thus the existence of a particular result record in a particular ORAM is hidden.

	\subsection{Practical improvements}\label{section:dp-improvements}

		Here we describe the optimizations aimed at bringing the construction's performance to the real-world demands.

		\subsubsection{ORAM request batching}\label{section:dp-improvements:oram-batching}

			We have noticed that although the entire set of ORAM requests for each query is known in advance, the requests are still executed sequentially.
			To address this inefficiency, we have designed a way to combine the requests in a batch and reduce the number of network requests to the bare minimum.
			We have implemented this method over PathORAM, which we use for the $(\eta_1, \eta_2)$-ORAM protocol, but the idea applies to most tree-based ORAMs (similar to~\cite{parallel-oram-improved}).

			Our optimization utilizes the fact that all PathORAM leaf IDs are known in advance and paths in a tree-based storage share the buckets close to the root.
			The core idea is to read all paths first, processes the requests and and then write all paths back.
			This way the client makes a single \texttt{read} request, which is executed much faster than many small requests.
			Requests are then processed in main memory, including re-encryptions.
			Finally, the client executes the \texttt{write} requests using remapped leaves as a single operation, saving again compared to sequential execution.

			This optimization provides up to \textbf{8 times} performance boost in our experiments.
			We note that the gains in speed and I/O overhead are achieved at the expense of main memory, which is not an issue given that the memory is released after a batch, and our experiments confirm that.
			The security guarantees of PathORAM are maintained with this optimization, since the security proof in \cite[Section 3.6]{path-oram} still holds. % chktex 2
			Randomized encryption, statistically independent remapping of leaves, and stash processing do not change.

		\subsubsection{Lightweight ORAM servers}\label{section:dp-improvements:three-tier}

			We have found in our experiments that na\"{\i}ve increase of the number of CPU cores and gigabytes of RAM does not translate into linear performance improvement after some threshold.
			Investigating the observation we have found that the \epsolute{} protocol, executing parallel ORAM protocols, is highly intensive with respect to main memory access, cryptographic operations and network usage.
			The bottleneck is the hardware --- we have confirmed that on a single machine the RAM and network are saturated quickly preventing the linear scaling.

			\begin{figure}[t]
	\centering
	\includegraphics[width=\linewidth]{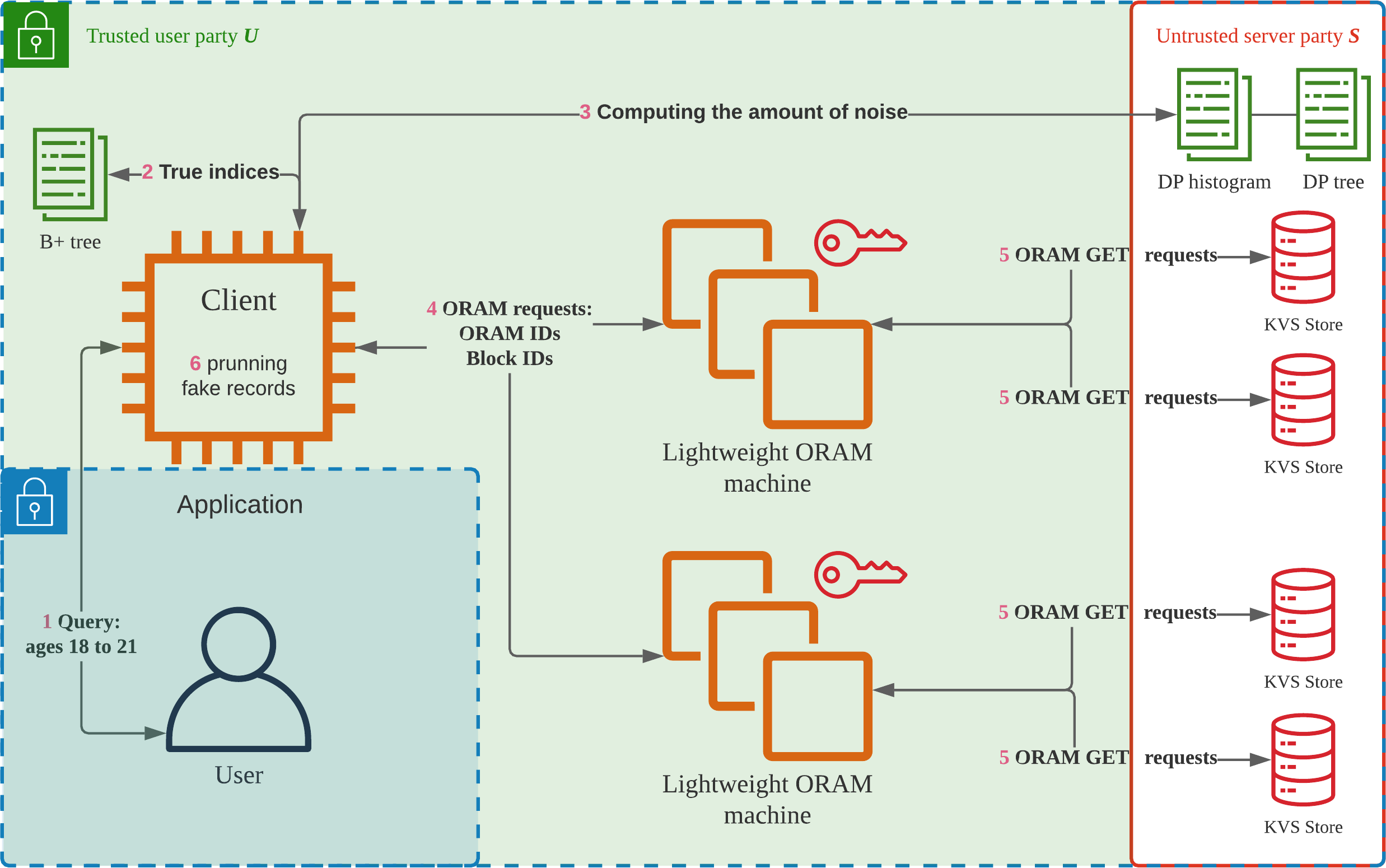}
	\Description{Lightweight ORAM machines diagram}
	\caption{
		Lightweight ORAM machines diagram.
		A \emph{user} sends a query to \user{} modeled as the \emph{client} machine, which uses local \emph{data index} and \emph{DP structures} to prepare a set of ORAM requests, which are sent to respective \emph{ORAM machines}.
		These machines execute the ORAM protocol against the \emph{untrusted storage} of \server{}.
	}%
	\label{figure:three-tier}
\end{figure}

			To address the problem, we split the user party \user{} into multiple lightweight machines that are connected locally to each other and reside in a single trust domain (e.g., same data center).
			Specifically, we maintain a \emph{client machine} that receives user requests and prepares ORAM \texttt{read} requests, and up to \oramsNumber{} lightweight \emph{ORAM machines}, whose only job is to run the ORAM protocols in parallel.
			See \cref{figure:three-tier} for the schematic representation of the architecture.
			We emphasize that \user{} is still a single party, therefore, the security and correctness guarantees remain valid.

			The benefit of this approach is that each of the lightweight machines has its own hardware stack.
			Communication overhead among \user{} machines is negligible compared to the one between \user{} and \server{}.
			The approach is also flexible: it is possible to use up to \oramsNumber{} ORAM machines and the machines do not have to be identical.
			Our experiments show that when the same number of CPU cores and amount of RAM are consumed the efficiency gain is up to \textbf{5 times}.

	\section{Experimental Evaluation}\label{section:experiments}

	We have implemented our solution as a modular client-server application in C++.
	We open-sourced all components of the software set: PathORAM~\cite{github-path-oram} and B+ tree~\cite{github-b-plus-tree} implementations and the main query executor~\cite{github-epsolute}.
	We provide PathORAM and B+ tree components as C++ libraries to be used in other projects; the code is documented, benchmarked and tested (228 tests covering \SI{100}{\percent} of the code).
	We have also published our datasets and query sets~\cite{our-datasets}.

	For cryptographic primitives, we used OpenSSL library (version 1.1.1i).
	For symmetric encryption in ORAM we have used AES-CBC algorithm~\cite{nist-aes,nist-modes} with a 256-bits key (i.e., $\eta_2 = 2^{-256}$), for the hash algorithm \algo{H} used to partition records among ORAMs we have used SHA-256 algorithm~\cite{nist-hash}.
	Aggregate tree fanout \fanout{} is 16, proven to be optimal in~\cite{hierarchical-methods-for-dp}.

	We designed our experiments to answer the following questions:
	\newlength{\questionLength}
	\settowidth{\questionLength}{Question-5}
	\begin{description}[
		font=\bfseries,
		leftmargin=\dimexpr\questionLength+1.0em\relax,
		labelindent=0pt,
		labelwidth=\questionLength%
	]
		\item[Question-1\label{item:question-practicality}] How practical is our system compared to the most efficient and most private real-world solutions?
		\item[Question-2\label{item:question-storage}] How practical is the storage overhead?
		\item[Question-3\label{item:question-parameters}] How different inputs and parameters of the system affect its performance?
		\item[Question-4\label{item:question-scalability}] How well does the system scale?
		\item[Question-5\label{item:question-optimizations}] What improvements do our optimizations provide?
		\item[Question-6\label{item:question-attributes}] What is the impact of supporting multiple attributes?
	\end{description}

	To address \ref{item:question-practicality} we have run the default setting using conventional RDBMS (MySQL and PostgreSQL), Linear Scan approach and Shrinkwrap~\cite{shrinkwrap}. % chktex 2
	To target \ref{item:question-storage}, we measured the exact storage used by the client and the server for different data, record and domain sizes. % chktex 2
	To answer \ref{item:question-parameters}, we ran a default setting and then varied all parameters and inputs, one at a time. % chktex 2
	For \ref{item:question-scalability} we gradually added vCPUs, ORAM servers and KVS instances and observed the rate of improvement in performance. % chktex 2
	For \ref{item:question-optimizations} we have run the default setting with our optimizations toggled. % chktex 2
	Lastly, for \ref{item:question-attributes} we have used two datasets to construct two indices and then queried each of the attributes. % chktex 2

	\subsection{Data sets}\label{section:experiments:data-sets}

		We used two real and one synthetic datasets --- California public pay pension database 2019~\cite{ca-employees-dataset} (referred to as ``CA employees''), Public Use Microdata Sample from US Census 2018~\cite{pums-dataset} (referred to as ``PUMS'') and synthetic uniform dataset.
		We have used salary / wages columns of the real datasets, and the numbers in the uniform set also represent salaries.
		The \texttt{NULL} and empty values were dropped.

		We created three versions of each dataset --- $10^5$, $10^6$ and $10^7$ records each.
		For uniform dataset, we simply generated the target number of entries.
		For PUMS dataset, we picked the states whose number of records most closely matches the target sizes (Louisiana for $10^5$, California for $10^6$ and the entire US for $10^7$).
		Uniform dataset was also generated for different domain sizes --- number of distinct values for the record.
		For CA employees dataset, the set contains \num{260 277} records, so we contracted it and expanded in the following way.
		For contraction we uniformly randomly sampled $10^5$ records.
		For expansion, we computed the histogram of the original dataset and sampled values uniformly within the bins.

		Each of the datasets has a number of corresponding query sets.
		Each query set has a selectivity or range size, and is sampled either uniformly or following the dataset distribution (using its CDF).

	\subsection{Default setting}\label{section:experiments:default-setting}

		The default setting uses the \protocolGamma{} from \cref{section:prallel-dp-oram} and lightweight ORAM machines from \cref{section:dp-improvements:three-tier,figure:three-tier}.
		We choose the \protocolGamma{} because it outperforms \protocolNoGamma{} in all experiments (see \textbf{\ref{item:question-scalability}} in \cref{section:experiments:results}).
		In the setting, there are 64 Redis services (8 services per one Redis server VM), 8 ORAM machines communicating with 8 Redis services each, and the client, which communicates with these 8 ORAM machines.
		We have empirically found this configuration optimal for the compute nodes and network that we used in the experiments.
		ORAM and Redis servers run on GCP \texttt{n1-standard-16} VMs (Ubuntu 18.04), in regions \texttt{us-east4} and \texttt{us-east1} respectively.
		Client machine runs \texttt{n1-highmem-16} VM in the same region as ORAM machines.
		The ping time between the regions (i.e.\ between trusted and untrusted zones) is \SI{12}{\milli\second} and the effective bandwidth is \SI{150}{\mega\byte\per\second}.
		Ping within a region is negligible.

		Default DP parameters are $\epsilon = \ln(2) \approx \num{0.693}$ and $\beta = 2^{-20}$, which are consistent with the other DP applications proposed in the literature~\cite{choosing-epsilon}.
		Buckets number is set as the largest power of $\fanout = 16$ that is no greater than the domain of the dataset \domainSize{}.

		Default dataset is a uniform dataset of $10^6$ records with domain size $10^4$, and uniformly sampled queries with selectivity \SI{0.5}{\percent}.
		Default record size is \SI{4}{\kibi\byte}.

	\subsection{Experiment stages}

		Each experiment includes running 100 queries such that the overhead is measured from loading query endpoints into memory to receiving the exact and whole query response from all ORAM machines.
		The output of an experiment is, among other things, the overhead (in milliseconds), the number of real and noisy records fetched and communication volume averaged per query.

	\subsection{RDBMS, Linear Scan and Shrinkwrap}

		On top of varying the parameters, we have run similar workloads using alternative mechanisms --- extremes representing highest performance or highest privacy.
		Unless stated otherwise, the client and the server are in the trusted and untrusted regions respectively, with the network configuration as in \cref{section:experiments:default-setting}.

		\subsubsection*{Relational databases}

			Conventional RDBMS represents the most efficient and least private and secure solution in our set.
			While MySQL and PostgreSQL offer some encryption options and no differential privacy, for our experiments we turned off security features for maximal performance.
			We have run queries against MySQL and PostgreSQL varying data and record sizes.
			We used \texttt{n1-standard-32} GCP VMs in \texttt{us-east1} region, running MySQL version 14.14 and PostgreSQL version 10.14.

		\subsubsection*{Linear Scan}

			Linear scan is a primitive mechanism that keeps all records encrypted on the server then downloads, decrypts and scans the entire database to answer every query.
			This method is trivially correct, private and secure, albeit not very efficient.
			There are RDBMS solutions, which, when configured for maximum privacy, exhibit linear scan behavior (e.g., MS-SQL Always Encrypted with Randomized Encryption~\cite{mssql-always-enc} and Oracle Column Transparent Data Encryption~\cite{oracle-tde}).
			For a fair comparison we make the linear scan even more efficient by allowing it to download data via parallel threads matching the number of threads and bytes per request to that of our solution.
			Although linear scan is wasteful in the amount of data it downloads and processes, compared to our solution it has a benefit of not executing an ORAM protocol with its logarithmic overhead and network communication in both directions.

		\subsubsection*{Shrinkwrap}

			Shrinkwrap~\cite{shrinkwrap} is a construction that answers federated SQL queries hiding both access pattern and communication volume.
			Using the EMP toolkit~\cite{emp-toolkit} and the code Shrinkwrap authors shared with us, we implemented a prototype that only answers range queries.
			This part of Shrinkwrap amounts to making a scan over the input marking the records satisfying the range, sorting the input, and then revealing the result set plus DP noise to the client.
			For the latter part we have adapted Shrinkwrap's Truncated Laplace Mechanism~\cite[Definition 4]{shrinkwrap} to hierarchical method~\cite{hierarchical-methods-for-dp} in order to be able to answer an unbounded number of all possible range queries.
			We have emulated the outsourced database setting by using two \texttt{n1-standard-32} servers in different regions (\SI{12}{\milli\second} ping and \SI{150}{\mega\byte\per\second} bandwidth) executing the algorithm in a circuit model (the faster option per Shrinkwrap experiments) and then revealing the result to the trusted client.
			We note that although the complexity of a Shrinkwrap query is $\bigO{n \log n}$ due to the sorting step, its functionality is richer as it supports more relational operators, like \texttt{JOIN}, \texttt{GROUP BY} and aggregation.
			We also note that since MySQL, PostgreSQL and Shrinkwrap are not parallelized within the query, experiments using more CPUs do not yield higher performance.

	\subsection{Results and Observations}\label{section:experiments:results}

		After running the experiments, we have made the following observations.
		Note that we report results based on the default setting.
		\begin{itemize}[leftmargin=*]
			\item
				\epsolute{} is efficient compared to a strawman approach, RDBMS and Shrinkwrap: it is three orders of magnitude faster than Shrinkwrap, 18 times faster than the scan and only 4--8 times slower than a conventional database.
				In fact, for different queries, datasets, and record sizes, our system is much faster than the linear scan, as we show next.
			\item
				\epsolute{}'s client storage requirements are very practical: client size is just below \SI{30}{\mega\byte} while the size of the offloaded data is over 400 times larger.
			\item
				\epsolute{} scales predictably with the change in its parameters: data size affects performance logarithmically, record size --- linearly, and privacy budget $\epsilon$ --- exponentially.
			\item
				\epsolute{} is scalable: using \protocolGamma{} with the lightweight ORAM machines, the increase in the number of threads translates into linear performance boost.
			\item
				The optimizations proposed in \cref{section:dp-improvements} provide up to an order of magnitude performance gain.
			\item
				\epsolute{} efficiently supports multiple indexed attributes.
				The overhead and the client storage increase slightly due to a lower privacy budget and extra local indices.
		\end{itemize}

		For the purposes of reproducibility we have put the log traces of all our experiments along with the instructions on how to run them on a publicly available page \href{https://epsolute.org}{epsolute.org}.
		Unless stated otherwise, the scale in the figures is linear and the $x$-axis is categorical.

		\subsubsection*{\textbf{\texorpdfstring{\ref{item:question-practicality}:}{} against RDBMS, Linear Scan and Shrinkwrap}}

			\begin{figure}[!ht]
	\centering
	\includegraphics[width=\columnwidth]{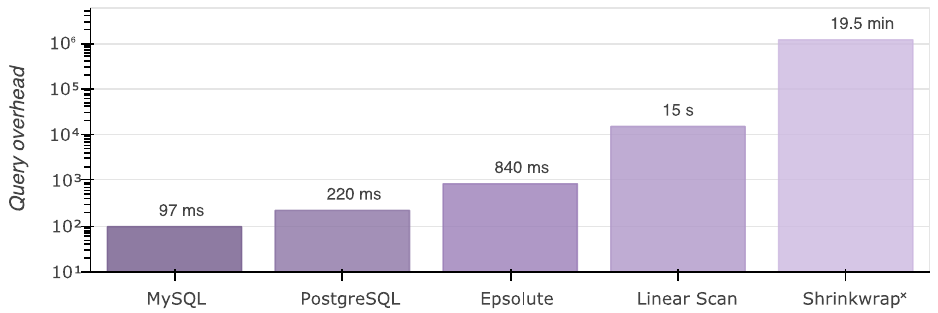}
	\Description{Different range-query mechanisms (log scale)}
	\caption{
		Different range-query mechanisms (log scale).
		Default setting: $10^6$ \SI{4}{\kibi\byte} uniformly-sampled records with the range $10^4$.
	}%
	\label{figure:mechanism}
\end{figure}

			The first experiment we have run using \epsolute{} is the default setting in which we observed the query overhead of \textbf{\SI[detect-all=true]{840}{\milli\second}}.
			To put this number in perspective, we compare \epsolute{} to conventional relational databases, the linear scan and Shrinkwrap.

			For the default setting, MySQL and PostgreSQL, configured for no privacy and maximum performance, complete in \SI{97}{\milli\second} and \SI{220}{\milli\second} respectively, which is just \textbf{8 to 4 times} faster than \epsolute{}, see \cref{figure:mechanism}.
			Conventional RDBMS uses efficient indices (B+ trees) to locate requested records and sends them over without noise and encryption, and it does so using less hardware resources.
			In our experiments RDBMS performance is linearly correlated with the result and record sizes.

			\begin{figure}[!ht]
	\centering
	\includegraphics[width=\columnwidth]{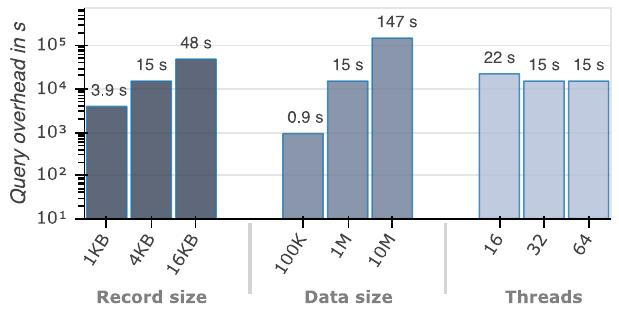}
	\Description{Linear scan performance, logarithmic scale}
	\caption{
		Linear scan performance, logarithmic scale.
		The experiments are run for the default setting of $10^6$ records of size \SI{4}{\kibi\byte} and 64 threads, with one of the three parameters varying.
	}%
	\label{figure:linear-scan}
\end{figure}

			Linear scan experiments demonstrate the practicality of \epsolute{} compared to a trivial ``download everything every time'' approach, see \cref{figure:linear-scan}.
			Linear scan's overhead is $\bigO{n}$ regardless of the queries, while \epsolute{}'s overhead is $\bigO{\log{n}}$ times the result size.
			According to our experiments, \epsolute{} eclipses the linear scan at \SI{4}{\kibi\byte}, 64 threads and only \emph{ten thousand records} (both mechanisms complete in about \SI{120}{\milli\second}).
			For a default setting (at a million records), the difference is \textbf{18 times}, see \cref{figure:linear-scan}.

			Because Shrinkwrap sorts the input obliviously in a circuit model, it incurs $\bigO{n \log n}$ comparisons, each resulting in multiple circuit gates, which is much more expensive than the linear scan.
			Unlike linear scan, however, Shrinkwrap does not require much client memory as the client merely coordinates the query.
			While Shrinkwrap supports richer set of relational operators, for range queries alone \epsolute{} is \textbf{three orders of magnitude} faster.

		\subsubsection*{\textbf{\texorpdfstring{\ref{item:question-storage}:}{} storage}}

			% chktex-file 26
% chktex-file 8

\newcommand{\lightvbar}{\color{lightGrey}\vrule}
\newcommand{\lightcline}[1]{\arrayrulecolor{lightGrey}\cline{#1}\arrayrulecolor{black}}

\begin{table}[!ht]
	\begin{adjustbox}{width=\linewidth}
		\sisetup{detect-all = true}
		\begin{tabular}{c | r !{\lightvbar{}} r | r !{\lightvbar{}} r | r !{\lightvbar{}} r} % chktex 44
			\toprule
				\diagbox{\dataSize}{\scriptsize{Record}}			& \multicolumn{2}{c !{\lightvbar{}}}{\SI{1}{\kibi\byte}}				& \multicolumn{2}{c !{\lightvbar{}}}{\SI{4}{\kibi\byte}}				& \multicolumn{2}{c}{\SI{16}{\kibi\byte}}												\\
			\midrule
				\multirow{2}{*}{$10^5$}								& \small\SI{400}{\kibi\byte}			& \small\SI{400}{\byte}			& \small\SI{400}{\kibi\byte}			& \small\SI{102}{\kibi\byte}	& \small\SI{400}{\kibi\byte}					& \small\SI{1.6}{\mega\byte}			\\ \lightcline{2-3} \lightcline{4-5} \lightcline{6-7}
																	& \small\bfseries\SI{396}{\mega\byte}	& \small\SI{4.6}{\mega\byte}	& \small\bfseries\SI{1.5}{\giga\byte}	& \small\SI{14}{\mega\byte}		& \small\bfseries\SI{6.2}{\giga\byte}			& \small\SI{51}{\mega\byte}				\\

			\midrule
				\multirow{2}{*}{$10^6$}								& \small\SI{3.9}{\mega\byte}			& \small\SI{400}{\byte}			& \small\SI{3.9}{\mega\byte}			& \small\SI{102}{\kibi\byte}	& \small\SI{3.9}{\mega\byte}					& \small\SI{1.6}{\mega\byte}			\\ \lightcline{2-3} \lightcline{4-5} \lightcline{6-7}
																	& \small\bfseries\SI{3.2}{\giga\byte}	& \small\SI{15}{\mega\byte}		& \small\bfseries\SI{12}{\giga\byte}	& \small\SI{25}{\mega\byte}		& \small\bfseries\SI{48}{\giga\byte}			& \small\SI{62}{\mega\byte}				\\

			\midrule
				\multirow{2}{*}{$10^7$}								& \small\SI{40}{\mega\byte}				& \small\SI{400}{\byte}			& \small\SI{40}{\mega\byte}				& \small\SI{102}{\kibi\byte}	& \small\itshape\SI{40}{\mega\byte}				& \small\itshape\SI{1.6}{\mega\byte}	\\ \lightcline{2-3} \lightcline{4-5} \lightcline{6-7}
																	& \small\bfseries\SI{24}{\giga\byte}	& \small\SI{99}{\mega\byte}		& \small\bfseries\SI{96}{\giga\byte}	& \small\SI{109}{\mega\byte}	& \small\itshape\bfseries\SI{384}{\giga\byte}	& \small\itshape\SI{146}{\mega\byte}	\\

			\midrule
				\diagbox[dir=SW, width=6em]{\dataSize}{\domainSize}	& \multicolumn{2}{c !{\lightvbar{}}}{$100$}								& \multicolumn{2}{c !{\lightvbar{}}}{$10^4$}							& \multicolumn{2}{c}{$10^6$}															\\
			\toprule
		\end{tabular}
		\sisetup{detect-none = true}
	\end{adjustbox}
	\caption{
		Storage usage for varying data, record and domain sizes.
		The values are as follows.
		Left top: index \indexI{} (B+ tree), right top: aggregate tree \serverDS{}, right bottom: ORAM \user{} state and left bottom (bold): ORAM \server{} state.
		\textit{Italic} indicates that the value is estimated.
	}%
	\label{table:storage}
\end{table}

			While \epsolute{} storage efficiency is near-optimal \efficiency{1}{0}, it is important to observe the absolute values.
			Index \indexI{} is implemented as a B+ tree with fanout 200 and occupancy \SI{70}{\percent}, and its size, therefore, is roughly $5.7 \dataSize$ bytes.
			Most of the ORAM client storage is the PathORAM stash with its size chosen in a way to bound failure probability to about $\eta_1 = 2^{-32}$ (see \cite[Theorem 1]{path-oram}). % chktex 2
			In \cref{table:storage}, we present \epsolute{} storage usage for the parameters that affect it --- data, record and domain sizes.
			We measured the sizes of the index \indexI{}, DP structure \serverDS{}, and ORAM client and server states.
			Our observations are:
			\begin{enumerate*}[label={(\roman*)}]
				\item index size expectedly grows only with the data size,
				\item \serverDS{} is negligibly small in practice,
				\item small \indexI{} and \serverDS{} sizes imply the efficiency of supporting multiple indexed attributes,
				\item \server{} to \user{} storage size ratio varies from \textbf{85} in the smallest setting to more than \textbf{\num[detect-all=true]{2000}} in the largest, and
				\item one can trade client storage for ORAM failure probability.
			\end{enumerate*}
			We conclude that the storage requirements of \epsolute{} are practical.

			% Stash size calculations are based on stash size 49, which, according to PathORAM paper, results in 14*0.6^{49} = 2^{-32} probability of failure

			% Total ORAM client size (in MB) is (( (2^h * 3 +3)*4 ) + (49 * (r * 1024) + 16))*64 / 1024^2
			% where h is ORAM_LOG_CAPACITY (11, 14, 17) and r is the record size in the number of kilobytes (1, 4, 16)

		\subsubsection*{\textbf{\texorpdfstring{\ref{item:question-parameters}:}{} varying parameters}}

			\begin{figure}[!ht]
	\centering
	\begin{minipage}{0.5\columnwidth}
		\centering
		\includegraphics[width=\textwidth]{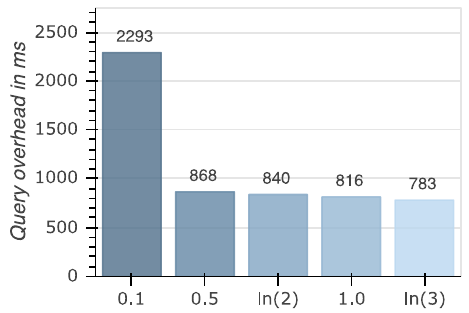}
		\Description{Privacy budget $\epsilon$}
		\captionof{figure}{Privacy budget $\epsilon$}%
		\label{figure:epsilon}
	\end{minipage}
	~ % chktex 39
	\begin{minipage}{0.5\columnwidth}
		\centering
		\includegraphics[width=\textwidth]{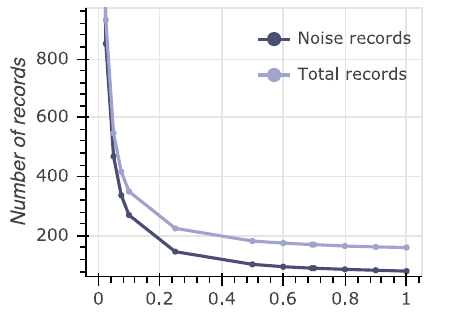}
		\Description{Effect of $\epsilon$}
		\captionof{figure}{Effect of $\epsilon$}%
		\label{figure:epsilon-effect}
	\end{minipage}
\end{figure}

			To measure and understand the impact of configuration parameters on the performance of our solution we have varied $\epsilon$, record size, data size \dataSize{}, domain size \domainSize{}, selectivities, as well as data and query distributions.
			The relation that is persistent throughout the experiments is that for given data and record sizes, the performance (the time to completely execute a query) is strictly proportional to the total number of records, fake and real, that are being accessed per query.
			Each record access goes through the ORAM protocol, which, in turn, downloads, re-encrypts and uploads $\bigO{\log{\dataSize}}$ blocks.
			These accesses contribute the most to the overhead and all other stages (e.g., traversing index or aggregate tree) are negligible.

			\paragraph*{Privacy budget \texorpdfstring{$\epsilon$}{epsilon} and its effect}

				We have run the default setting for $\epsilon = \{ 0.1, \allowbreak 0.5, \allowbreak \ln{2}, \allowbreak 1.0, \allowbreak \ln{3} \}$.
				$\epsilon$ strictly contributes to the amount of noise, which grows exponentially as $\epsilon$ decreases, see \cref{figure:epsilon}, observe sharp drop.
				As visualized on \cref{figure:epsilon-effect}, at high $\epsilon$ values the noise contributes a fraction of total overhead, while at low values the noise dominates the overhead entirely.

			\begin{figure}[!ht]
	\centering
	\includegraphics[width=\columnwidth]{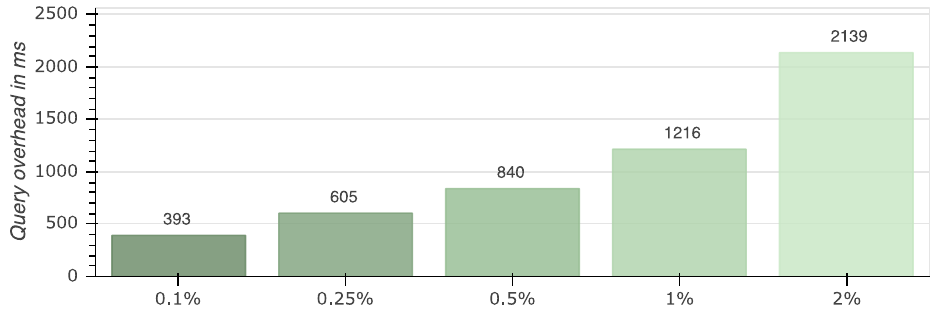}
	\Description{Selectivity}
	\caption{Selectivity}%
	\label{figure:selectivity}
\end{figure}

			\paragraph*{Selectivity}

				We have ranged the selectivity from \SI{0.1}{\percent} to \SI{2}{\percent} of the total number of records, see \cref{figure:selectivity}.
				Overhead expectedly grows with the result size.
				For smaller queries, and thus for lower overhead, the relation is positive, but not strictly proportional.
				This phenomena, observed for the experiments with low resulting per-query time, is explained by the variance among parallel threads.
				During each query the work is parallelized over \oramsNumber{} ORAMs and the query is completed when the \emph{last} thread finishes.
				The problem, in distributed systems known as ``the curse of the last reducer''~\cite{curse-of-last-reducer}, is when one thread takes disproportionally long to finish.
				In our case, we run 64 threads in default setting, and the delay is usually caused by a variety of factors --- blocking I/O, network delay or something else running on a shared vCPU\@.
				This effect is noticeable when a single thread does relatively little work and small disruptions actually matter; the effect is negligible for large queries.

			\begin{figure}[!ht]
	\centering
	\begin{minipage}{0.31\columnwidth}
		\centering
		\includegraphics[width=\textwidth]{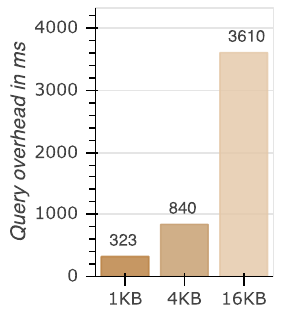}
		\Description{Record size}
		\captionof{figure}{Record size}%
		\label{figure:record-size}
	\end{minipage}
	~ % chktex 39
	\begin{minipage}{0.31\columnwidth}
		\centering
		\includegraphics[width=\textwidth]{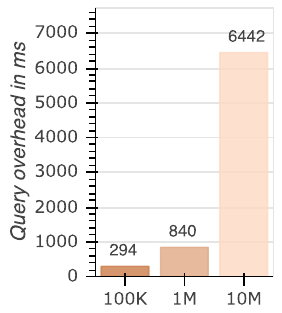}
		\Description{Data size}
		\captionof{figure}{Data size}%
		\label{figure:data-size}
	\end{minipage}
	~ % chktex 39
	\begin{minipage}{0.38\columnwidth}
		\centering
		\includegraphics[width=0.816\textwidth]{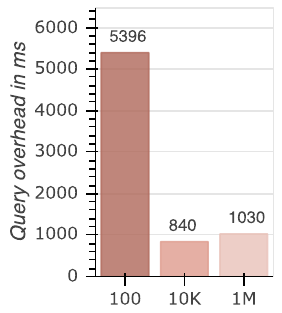} % 0.31/0.38
		\Description{Domain size}
		\captionof{figure}{Domain size}%
		\label{figure:domain-size}
	\end{minipage}
\end{figure}

			\paragraph*{Record, data and domain sizes}

				We have tried \SI{1}{\kibi\byte}, \SI{4}{\kibi\byte} and \SI{16}{\kibi\byte} records, see \cref{figure:record-size}.
				Trivially, the elapsed time is directly proportional to the record size.

				We set \dataSize{} to $10^5$, $10^6$ and $10^7$, see \cref{figure:data-size}.
				The observed correlation of overhead against the data size is positive but non-linear, 10 times increment in \dataSize{} results in less than 10 times increase in time.
				This is explained by the ORAM overhead --- when \dataSize{} changes, the ORAM storage gets bigger and its overhead is logarithmic.

				For synthetic datasets we have set \domainSize{} to $100$, $10^4$ and $10^6$, see \cref{figure:domain-size}.
				The results for domain size correlation are more interesting: low and high values deliver worse performance than the middle value.
				Small domain for a large data set means that a query often results in a high number of real records, which implies significant latency regardless of noise parameters.
				A sparse dataset, on the other hand, means that for a given selectivity wider domain is covered per query, resulting in more nodes in the aggregate tree contributing to the total noise value.

			\begin{figure}[!ht]
	\centering
	\begin{minipage}{0.5\columnwidth}
		\centering
		\includegraphics[width=0.75\textwidth]{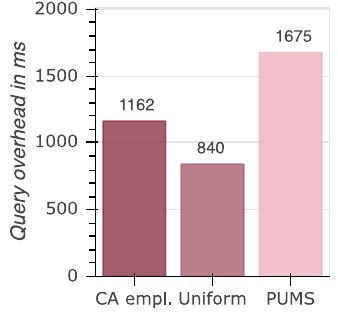}
		\Description{Data distribution}
		\captionof{figure}{Data distribution}%
		\label{figure:data-distribution}
	\end{minipage}
	~ % chktex 39
	\begin{minipage}{0.5\columnwidth}
		\centering
		\includegraphics[width=0.75\textwidth]{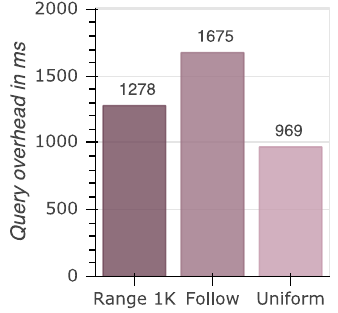}
		\Description{Query distribution}
		\captionof{figure}{Query distribution}%
		\label{figure:query-distribution}
	\end{minipage}
\end{figure}

			\paragraph*{Data and query distributions}

				Our solution performs best on the uniform data and uniform ranges, see \cref{figure:data-distribution,figure:query-distribution}.
				Once a skew of any kind is introduced, there appear sparse and dense regions that contribute more overhead than uniform regions.
				Sparse regions span over wider range for a given selectivity, which results in more noise.
				Dense regions are likely to include more records for a given range size, which again results in more fetched records.
				Both real datasets are heavily skewed towards smaller values as few people have ultra-high salaries.

		\subsubsection*{\textbf{\texorpdfstring{\ref{item:question-scalability}:}{} scalability}}

			\begin{figure}[!ht]
	\centering
	\includegraphics[width=\columnwidth]{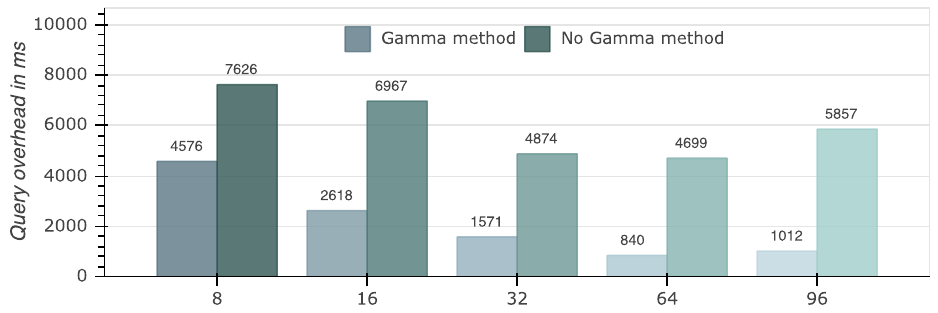}
	\Description{Scalability measurements for \protocolGamma{} and \protocolNoGamma{}}
	\caption{Scalability measurements for \protocolGamma{} and \protocolNoGamma{}}%
	\label{figure:scalability}
\end{figure}

			Horizontal scaling is a necessity for a practical system, this is the motivation for the parallelization in the first place.
			Ideally, performance should improve proportionally to the parallelization factor, number of ORAMs in our case, \oramsNumber{}.

			For scalability experiments we run the default setting for both \protocolNoGamma{} and \protocolGamma{} (\emph{no-$\gamma$-method} and \emph{$\gamma$-method} respectively) varying the number of ORAMs \oramsNumber{}, from 8 to 96 (maximum vCPUs on a GCP VM).
			The results are visualized on \cref{figure:scalability}.
			We report two positive observations:
			\begin{enumerate*}[label={(\roman*)}]
				\item the $\gamma$-method provides substantially better performance and storage efficiency, and
				\item when using this method the system scales linearly with the number of ORAMs.
			\end{enumerate*}
			($\oramsNumber = 96$ is a special case because some ORAMs had to share a single KVS.)

		\subsubsection*{\textbf{\texorpdfstring{\ref{item:question-optimizations}:}{} optimizations benefits}}

			\begin{table}[!ht]
	\begin{adjustbox}{width=\linewidth}
		\begin{tabular}{ l c c >{\bfseries}c }
			\toprule
				Improvement (section)													& Enabled					& Disabled					& Boost			\\
			\midrule
				ORAM batching (\ref{section:dp-improvements:oram-batching})				& \SI{840}{\milli\second}	& \SI{6978}{\milli\second}	& 8.3x			\\
				Lightweight ORAM machines (\ref{section:dp-improvements:three-tier})	& \SI{840}{\milli\second}	& \SI{4484}{\milli\second}	& 5.3x			\\
				Both improvements														& \SI{840}{\milli\second}	& \SI{8417}{\milli\second}	& \emph{10.0x}	\\
			\bottomrule
		\end{tabular}
	\end{adjustbox}
	\caption{Improvements over parallel \epsolute{}}%
	\label{table:optimizations}
\end{table}

			\cref{table:optimizations} demonstrates the boosts our improvements provide; when combined, the speedup is up to an order of magnitude.

			ORAM request batching (\cref{section:dp-improvements:oram-batching}) makes the biggest difference.
			We have run the default setting with and without the batching.
			The overhead is substantially smaller because far fewer I/O requests are being made, which implies benefits across the full stack: download, re-encryption and upload.

			Using lightweight ORAM machines (\cref{section:dp-improvements:three-tier}) makes a difference when scaling.
			In the default setting, 64 parallel threads quickly saturate the memory access and network channel, while spreading computation among nodes removes the bottleneck.

		\subsubsection*{\textbf{\texorpdfstring{\ref{item:question-attributes}:}{} multiple attributes}}

			\begin{figure}[!ht]
	\centering
	\includegraphics[width=\columnwidth]{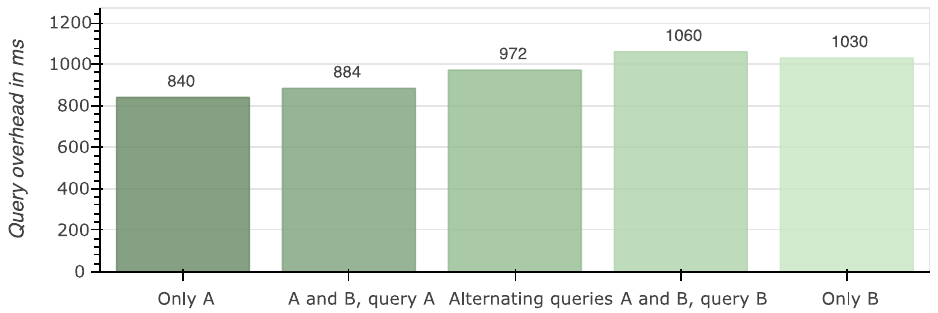}
	\Description{Query overhead when using multiple attributes}
	\caption{
		Query overhead when using multiple attributes.
		\emph{Only A} and \emph{Only B} index one attribute.
		\emph{A and B} indexes both attributes and then queries one of them.
		\emph{Alternating} indexes both attributes and runs half of the queries against \emph{A} and another half against \emph{B}.
	}%
	\label{figure:attributes}
\end{figure}

			\epsolute{} supports multiple indexed attributes.
			In \cref{section:dp-oram:multiple-attributes} we described that the performance implications amount to having an index \indexI{} and a DP structure \serverDS{} per attribute and sharing the privacy budget $\epsilon$ among all attributes.
			As shown in \cref{table:storage}, \indexI{} and \serverDS{} are the smallest components of the client storage.
			To observe the query performance impact, we have used the default dataset with domains $10^4$ and $10^6$ as indexed attributes \emph{A} and \emph{B} respectively.
			We ran queries against only \emph{A}, only \emph{B} and against both attributes in alternating fashion.
			Each of the attributes used $\epsilon = \frac{\ln{2}}{2}$ to match the default privacy budget of $\ln(2)$.

			\cref{figure:attributes} demonstrates the query overhead of supporting multiple attributes.
			The principal observation is that the overhead increases only slightly due to a lower privacy budget.
			The client storage went up by just \SI{9}{\mega\byte}, and still constitutes only \SI{3.3}{\percent} of the server storage, which is not affected by the number of indexed attributes.

			% for 10K domain: total client storage went from 3.9+0.1+25 to 3.9+0.1+25+9, i.e. 31%
			% for 1M domain: total client storage went from 3.9+1.6+25 to 3.9+1.6+25+9, i.e. 29.5%

	\section{Conclusion and Future Work}

	In this paper, we present a system called \epsolute{} that can be used to store and retrieve encrypted records in the cloud while providing strong and provable security guarantees, and that exhibits excellent query performance for range and point queries.
	We use an optimized Oblivious RAM protocol that has been parallelized together with very efficient Differentially Private sanitizers that hide both the access patterns and the exact communication volume sizes and can withstand advanced attacks that have been recently developed.
	We provide a prototype of the system and present an extensive evaluation over very large and diverse datasets and workloads that show excellent performance for the given security guarantees.

	In our future work, we plan to investigate methods to extend our approaches to use a trusted execution environment (TEE), like SGX, in order to improve the performance even further.
	We will also explore a multi-user setting without the need for a shared stateful client, and enabling dynamic workloads with insertions and updates.
	We will also consider how adaptive and non-adaptive security models would change in the case of dynamic environments.
	One would presumably also require DP of the server's view in this setting.
	Lastly, we plan to explore other relational operations like \texttt{JOIN} and \texttt{GROUP BY}.

	\begin{acks}

	We thank anonymous reviewers and Arkady Yerukhimovich for valuable feedback.
	We also thank Daria Bogatova for devising the name \epsolute{} and helping with the plots, diagrams and writing.
	Finally, we thank Johes Bater for sharing Shrinkwrap code and reviewing the prototype.
	Kobbi Nissim was supported by NSF Grant No.\ 2001041, ``Rethinking Access Pattern Privacy: From Theory to Practice''.
	Dmytro Bogatov and George Kollios were supported by NSF CNS-2001075 Award.

\end{acks}

	\bibliographystyle{style/ACM-Reference-Format}
	\balance%
	\bibliography{bibfile}

\end{document}